\documentclass[fleqn,usenatbib,twocolumn]{mnras}

\usepackage{newtxtext,newtxmath}
\usepackage{xspace}

\usepackage[T1]{fontenc}

\DeclareRobustCommand{\VAN}[3]{#2}
\let\VANthebibliography\thebibliography
\def\thebibliography{\DeclareRobustCommand{\VAN}[3]{##3}\VANthebibliography}

\usepackage{graphicx}	
\usepackage{amsmath}	
\usepackage{gensymb}
\usepackage{subcaption}
\usepackage{orcidlink}
\usepackage{booktabs}
\usepackage{siunitx}
\usepackage{anyfontsize} 
\usepackage[table]{xcolor}
\usepackage{upgreek}

\newcommand{\rone}{FRB~20121102A\xspace}
\newcommand{\ronetwin}{FRB~20190520B\xspace}

\newcommand{\rsixseven}{FRB~20201124A\xspace}
\newcommand{\roneoneseven}{FRB~20220912A\xspace}
\newcommand{\ronefourseven}{FRB~20240114A\xspace}
\newcommand{\ronezeronine}{FRB~20220529A\xspace}

\newcommand{\rthree}{FRB~20180916B\xspace}
\newcommand{\meightone}{FRB~20200120E\xspace}
\newcommand{\rmkt}{FRB~20240619D\xspace}
\newcommand{\reighteen}{FRB~20190417A\xspace}

\newcommand{\sfxc}{{\tt SFXC}\xspace}

\newcommand{\psrchive}{{\tt PSRCHIVE}\xspace}

\newcommand{\presto}{{\tt PRESTO}\xspace}

\def\torun{Toru\'n\xspace}
\def\nancay{Nan\c{c}ay\xspace}
\newcommand{\dmunit}{pc\,cm$^{-3}$\xspace}
\newcommand{\dmunityr}{pc\,cm$^{-3}$\,yr$^{-1}$\xspace}
\newcommand{\rmunit}{rad\,m$^{-2}$\xspace}
\newcommand{\rom}[1]{\uppercase\expandafter{\romannumeral #1\relax}}

\makeatother
\usepackage[switch]{lineno} 

\newcommand{\mktloc}{$\alpha=19^{\mathrm{h}}49^{\mathrm{m}}29.21^{\mathrm{s}}$, $\delta = -25\degr12\arcmin49.40\arcsec$ (J2000)\,}
\newcommand{\mktlocatel}{$\alpha=19^{\mathrm{h}}49^{\mathrm{m}}29.21^{\mathrm{s}}$, $\delta = -25\degr12\arcmin49.40\arcsec$ (J2000)\,}

\title[\rmkt: Local environment \& energetics]{A HyperFlash and \'ECLAT view of the local environment and energetics of the repeating \rmkt}

\author[O.S. Ould-Boukattine et al.]
{O.~S.~Ould-Boukattine \orcidlink{0000-0001-9381-8466}$^{1,2}$\thanks{E-mail: ouldboukattine@astron.nl},
A.~J.~Cooper \orcidlink{0000-0002-4033-3139}$^{3}$,
J.~W.~T.~Hessels \orcidlink{0000-0003-2317-1446}$^{4,5,2,1}$,
D.~M.~Hewitt \orcidlink{0000-0002-5794-2360}$^{2}$,
S.~K.~Ocker \orcidlink{0000-0002-4941-5333}$^{6,7}$, \newauthor
A.~Moroianu \orcidlink{0000-0003-1936-9062}$^{2}$, 
K.~Nimmo \orcidlink{0000-0003-0510-0740}$^{8}$,
M.~P.~Snelders\orcidlink{0000-0001-6170-2282}$^{1,2}$, 
I.~Cognard \orcidlink{0000-0002-1775-9692}$^{9,10}$,
T.~J.~Dijkema \orcidlink{0000-0001-7551-4493}$^{1,11}$, 
M.~Fine \orcidlink{0009-0006-1258-4228}$^{4,5,2}$,\newauthor
M.~P.~Gawro\'nski \orcidlink{0000-0002-5794-2360}$^{12}$, 
W.~Herrmann \orcidlink{0000-0001-5806-446X}$^{13}$,
J.~Huang \orcidlink{0000-0002-5794-2360}$^{4,5}$,
F.~Kirsten \orcidlink{0000-0002-5794-2360}$^{14,1}$, 
Z.~Pleunis \orcidlink{0000-0002-4795-697X}$^{2,1}$,
W.~Puchalska \orcidlink{0000-0003-2422-6605}$^{12}$,\newauthor
S.~Ranguin \orcidlink{0009-0004-1397-9331}$^{2}$ 
and T.~Telkamp \orcidlink{0009-0002-5048-2573}$^{11}$ \newauthor
\\
$^{1}$ASTRON, Netherlands Institute for Radio Astronomy, Oude Hoogeveensedijk 4, 7991 PD Dwingeloo, The Netherlands\\
$^{2}$Anton Pannekoek Institute for Astronomy, University of Amsterdam, Science Park 904, 1098 XH, Amsterdam, The Netherlands\\
$^{3}$Astrophysics, The University of Oxford, Keble Road, Oxford, OX1 3RH, UK\\
$^{4}$Trottier Space Institute, McGill University, 3550 rue University, Montr\'eal, QC H3A~2A7, Canada\\
$^{5}$Department of Physics, McGill University, 3600 rue University, Montr\'eal, QC H3A~2T8, Canada\\
$^{6}$Cahill Center for Astronomy and Astrophysics, California Institute of Technology, Pasadena, CA, USA\\
$^{7}$The Observatories of the Carnegie Institution for Science, Pasadena, CA, USA\\
$^{8}$MIT Kavli Institute for Astrophysics and Space Research, Massachusetts Institute of Technology, 77 Massachusetts Ave, Cambridge, MA 02139, USA\\
$^{9}$Station de Radioastronomie de Nan\c{c}ay, Observatoire de Paris, PSL University, CNRS, Universit\'{e} d'Orl\'{e}ans, F-18330 Nan\c{c}ay, France\\
$^{10}$Laboratoire de Physique et Chimie de l'Environnement et de l'Espace LPC2E UMR7328, Universit\'{e} d'Orl\'{e}ans, CNRS, F-45071 Orl\'{e}ans, France\\
$^{11}$CAMRAS Dwingeloo Radio Telescope Foundation, Oude Hoogeveensedijk 4, 7991 PD Dwingeloo, The Netherlands \\
$^{12}$Institute of Astronomy, Faculty of Physics, Astronomy and Informatics, Nicolaus Copernicus University, Grudziadzka 5, 87-100 \torun, Poland\\
$^{13}$Astropeiler Stockert e.V., Astropeiler 1-4, 53902 Bad M\"{u}nstereifel, Germany\\
$^{14}$Department of Space, Earth and Environment, Chalmers University of Technology, Onsala Space Observatory, 439 92, Onsala, Sweden\\
}

\date{Accepted 2026 January 6. Received 2025 December 9; in original form 2025 September 19}

\pubyear{\the\year{}}

\begin{document}
\label{firstpage}
\pagerange{\pageref{firstpage}--\pageref{lastpage}}
\maketitle

\begin{abstract}
Time-variable propagation effects provide a window into the local plasma environments of repeating fast radio burst (FRB) sources.
Here we report high-cadence observations of \rmkt, as part of the HyperFlash and \'ECLAT programs. We observed for $500$\,h and detected $217$~bursts, including 10 bursts with high fluence ($>25$\,Jy~ms) and implied energy.
We track burst-to-burst variations in dispersion measure (DM) and rotation measure (RM), from which we constrain the parallel magnetic field strength in the source's local environment: 
$0.27\pm0.13$\,mG.
Apparent DM variations between sub-bursts in a single bright event are interpreted as coming from plasma lensing or variable emission height.
We also identify two distinct scintillation screens along the line of sight, one associated with the Milky Way and the other likely located in the FRB's host galaxy or local environment.
Together, these (time-variable) propagation effects reveal that \rmkt is embedded in a dense, turbulent and highly magnetised plasma. The source's environment is more dynamic than that measured for many other (repeating) FRB sources, but less extreme compared to several repeaters that are associated with a compact, persistent radio source.
\rmkt's cumulative burst fluence distribution shows a power-law break, with a flat tail at high energies. Along with previous studies, this emphasises a common feature in the burst energy distribution of hyperactive repeaters.
Using the break in the burst fluence distribution, we estimate a source redshift of $z=0.042$-$0.240$.
We discuss \rmkt's nature in the context of similar studies of other repeating FRBs. 
\end{abstract}


\clearpage

\section{Introduction}
Out of the thousands of fast radio burst (FRB) sources detected to date, about a half dozen are prolific repeaters that have been detected hundreds to thousands of times by various telescopes \citep[e.g.,][]{chime_2021_apjs, xu_2022_natur, konijn_2024_mnras}. When active, these `hyperactive' repeaters may even contribute a significant fraction of the total observable all-sky rate of FRBs above a certain fluence limit ($22.0^{+15.6}_{-10.3}\,\%$\,$\mathrm{R_{sky}}\,(\mathcal{F}>500\mathrm{~Jy~ms})$, e.g., \citealt{ouldboukattine_2025_mnras}). The majority of repeaters have repetition rates at least an order of magnitude lower \citep[Fig.~9]{chime_2023_apj}, and are roughly consistent with the upper limits on the rates of most apparent non-repeaters, making it difficult to establish two clearly separated types of FRBs \citep{chime_2023_apj, kirsten_2024_natas}. Though rare, hyperactive repeaters are valuable because they allow one to build a record of an FRB's activity as a function of radio frequency and time \citep[e.g.,][]{hewitt_2022_mnras, jahns_2023_MNRAS}. By measuring propagation effects like dispersion \citep{hessels_2019_apjl}, Faraday rotation \citep{gopinath_2024_mnras}, and scintillation/scattering \citep{nimmo_2025_natur,ocker_2023_mnras}, we can also map dynamic effects along the line of sight and in the FRB's local environment. This information complements what we can glean from precision localisations within a host galaxy \citep{hewitt_2024_mnras, chime_2025_apjl, bhardwaj_2025_apjl}. Together, these are the clues needed to differentiate between proposed FRB models, and to ascertain whether a single model can explain all sources.
\par
Such studies are particularly valuable when full polarimetry, amplitude and phase data (`voltage'/`baseband' data) are available to flexibly study the bursts at a range of time and frequency resolutions. Exploring microsecond temporal structures can strongly constrain the dispersion measure (DM) and possible subtle variations \citep{nimmo_2022_natas}, while wide-band polarimetry can detect rotation measure (RM) variations \citep[e.g.,][]{li_2025_arxiv}.

Some repeating FRBs have been found in extreme and dynamic magneto-ionic environments \citep{michilli_2018_natur, annathomas_2023_sci, moroianu_2026_apj}. The first-known repeater, \rone, resides near an active star-forming region in low-metallicity dwarf host galaxy \citep{chatterjee_2017_nature, tendulkar_2017_apjl, bassa_2017_apjl}. It shows both a highly variable DM and RM, and has been associated with a persistent radio source \citep[PRS,][]{marcote_2017_apjl} that is possibly a nebula powered by the FRB source. \ronetwin is another repeater whose bursts and host galaxy are remarkably similar to \rone \citep{annathomas_2023_sci}. In contrast, the repeaters \rthree and \meightone have shown modest to no DM or RM variations and have no associated PRS down to stringent luminosity limits \citep{mckinven_2023_apj,nimmo_2021_natas}. Repeating FRBs are found in a diverse range of galaxies and local environments --- including globular clusters \citep{kirsten_2022_natur}, dwarf galaxies (\citealp{hewitt_2024_apjl}; \citealp{moroianu_2026_apj}; \citealt{bhardwaj_2025_apjl}), star-forming spirals \citep{marcote_2020_natur}, and ellipticals \citep{shah_2025_apjl, eftekhari_2025_apjl} --- raising the question of whether repeating FRBs share a common progenitor. Similarly, the connection between repeating FRBs and the dominant population of apparently non-repeating FRBs is currently unclear (see, e.g., \citealt{kirsten_2024_natas} and \citealt{chime_2025_apjl}).

Previous studies have found that repeating FRBs are typically longer in duration and narrower in emission bandwidth compared to the apparently one-off bursts \citep{pleunis_2021_apj}. Repeaters also often show a time-frequency drift, where subbursts typically appear at lower frequencies at later times (colloquially termed the `sad trombone' effect; \citealt{hessels_2019_apjl,faber_2024_apj}). This suggests a different emission mechanism and potentially progenitor type.  Nonetheless, the hyperactive sources are clearly exceptionally active, and distinguish themselves from much of the known population. Burst energetics are another valuable point of comparison. High-cadence observations of repeaters have shown that high-energy bursts have a similar statistical distribution to the population of one-off bursts and can reach similarly high energies --- implying that the source's total energy reservoir is comparable \citep{kirsten_2024_natas, ouldboukattine_2025_mnras}. This suggests a possible link, whereby the apparently one-off bursts may simply be the most energetic bursts from sources that will eventually repeat.

Recently, the MeerTRAP collaboration discovered a new hyperactive repeater, \rmkt 
using the MeerKAT telescope \citep{tian_2024_atel, tian_2025_mnras}. The source was revealed to be highly active due to the detection of $3$ bursts within only $2$\,minutes. \citet{tian_2024_atel} used their voltage data to localise \rmkt to \mktloc with an uncertainty of $0.9\arcsec$. Two optical sources were identified in the DESI Legacy Survey DR10 near the position of the FRB, but neither has a measured redshift \citep{dey_2019_aj} and the host galaxy of \rmkt\ remains unknown. Table~\ref{tab:rmkt-properties} provides an overview of the properties of \rmkt. Follow-up observations by \cite{kumar_2024_atel} and \cite{bhusare_2025_apj} using the upgraded Giant Metrewave Radio Telescope (uGMRT) led to the detection of $60$ bursts between $550$--$750$\,MHz in July 2024 and placed a 5-$\sigma$ upper limit on the association of a PRS of $417$\,$\upmu$Jy\,beam$^{-1}$ at $650$\,MHz. \rmkt provides a new opportunity to study a hyperactive repeater and to compare its energetics and local environment with the small sample of well-studied FRB sources.

In this paper, we report on a high-cadence observing campaign towards \rmkt, totalling more than $500$\,h of observations spanning over $4$ months and yielding $217$ burst detections including $10$ very-high-energy bursts (fluence $>25$\,Jy\,ms). These observations were carried out as part of the HyperFlash FRB monitoring campaign, with three $25$-m radio telescopes located in Europe: The Westerbork RT-1 (Wb) and Dwingeloo (Dw) telescopes,  both located in the Netherlands, and the Stockert telescope (St) located in Germany. Additionally, we also observed with the \nancay Radio Telescope (NRT) as part of the \'ECLAT observing program. We measure the spectrotemporal burst properties for our sample to investigate time variability. Additionally, we compare the cumulative burst fluence distribution with other repeating FRB sources in the literature. 

In Section~\ref{sec:obs_burst_search} we describe the data acquisition, observational set-up, and burst search. Section~\ref{sec:ana_res} presents the analysis of the burst properties and time-variable propagation effects. In Section~\ref{sec:discussion} we discuss our results and the summarise conclusions in Section~\ref{sec:conclusion}.

\section{Observations and burst search} \label{sec:obs_burst_search}

\begin{table}
\caption{Summary of known, measured, and inferred properties of \rmkt.}
\label{tab:rmkt-properties}
\resizebox{\columnwidth}{!}{%
\begin{tabular}{l c}
\toprule
\toprule
Property         & Values \\ \midrule
Right Ascension (J2000, ICRS)$^\mathrm{a}$                      & $19^{\mathrm{h}}49^{\mathrm{m}}29.21^{\mathrm{s}}$ \\
Declination (J2000, ICRS)$^\mathrm{a}$                          & $-25\degr12\arcmin49.40\arcsec$     \\
Uncertainty (RA \& DEC)$^\mathrm{a}$                            & $0.9\arcsec$ \\ 
Galactic longitude ($\ell$)                                     & $15.4559\degree$ \\ 
Galactic latitude ($b$)                                         & $-23.3294\degree$ \\ 
\midrule
Galactic scattering timescale ($1$\,GHz)$^\mathrm{b}$           & $0.73$\,$\upmu$s \\ 
Galactic scattering timescale ($1.271$\,GHz)$^\mathrm{b}$       & $0.28$\,$\upmu$s \\ 
Galactic scintillation bandwidth ($1$\,GHz)$^\mathrm{b}$        & $251.3$\,kHz \\ 
Galactic scintillation bandwidth ($1.271$\,GHz)$^\mathrm{b}$    & $655.7$\,kHz \\ 
$\textrm{DM}_\textrm{disk}$ (NE2001)$^\mathrm{b}$               & $94.3$\,\dmunit \\
$\textrm{DM}_\textrm{disk}$ (YMW16)$^\mathrm{c}$                & $62.7$\,\dmunit \\ 
\midrule
\midrule
\multicolumn{2}{l}{Measured and inferred properties in this work} \\ 
\midrule
Dispersion measure range                                        & $[464.857,\ 465.266]$\,\dmunit \\ 
Rotation measure range                                          & $[-182.2,\ -279.8]$\,\rmunit \\ 
Scattering timescale range ($1$\,GHz)                           & $[120,\,2900]$\,$\upmu$s$^\mathrm{d}$ \\
Galactic Scintillation bandwidth ($1.271$\,GHz)                 & $[74,\,159]$\,kHz \\ 
Extragalactic Scintillation bandwidth ($1.271$\,GHz)            & $6.7 \pm 0.7$\,kHz$^\mathrm{e}$ \\ 
Parallel magnetic field strength ($B_{\parallel}$)              & $0.27\pm0.13 \, {\rm mG}$ \\ 
Redshift ($z$) upper limit Macquart (DM-$z$) relation           & $<0.37$ \\ 
Redshift ($z$) estimate based on $\mathrm{E}_\mathrm{break}$    & $0.042-0.240$ \\ 
\bottomrule
\bottomrule
\multicolumn{2}{l}{$\mathrm{^{a}}$From the MeerKat disovery paper \citep{tian_2025_mnras}.} \\
\multicolumn{2}{l}{$\mathrm{^{b}}$Calculated using the NE2001p model \citep{cordes_2002_arxiv, ocker_2024_rnaas}.} \\
\multicolumn{2}{l}{$\mathrm{^{c}}$Calculated using the YMW16 model \citep{yao_2017_apj}.} \\
\multicolumn{2}{l}{$\mathrm{^{d}}$referenced to $1$\,GHz, see Section \ref{sec:scattering}.} \\
\multicolumn{2}{l}{$\mathrm{^{e}}$Based on a single measurement, B01-Wb, see Section \ref{sec:scint} and Figure \ref{fig:scint}.} \\
\end{tabular}%
}
\end{table}

We observed \rmkt using several European radio telescopes, accumulating $504.1$\,h on source between 5 July 2024 (MJD~$60496$) to 19 October 2024 (MJD~$60602$). Our first observation started only $13$\,h after the announcement of the source's discovery ATel \citep[][]{tian_2024_atel}, as shown in the overview of our observational campaign in Appendix Figure~\ref{fig:obs_campaign}. Appendix Table~\ref{tab:obs_coverage} further summarises the observational campaign, including the fluence detection limits of each telescope. 

\subsection{HyperFlash and \'ECLAT}

HyperFlash (PI: O.~S.~Ould-Boukattine) is a high-cadence monitoring program using 25-m diameter telescopes to observe repeating and potentially repeating FRB sources. With HyperFlash, we aim to observe the brightest and rarest FRB events by monitoring each source for hundreds to thousands of hours --- predominantly around $1.3$\,GHz, but occasionally around $350$\,MHz and $5$\,GHz central frequencies as well. Our HyperFlash campaign on \rmkt used the 25-m Westerbork RT-1 telescope (the Netherlands), the 25-m Stockert telescope (Germany), and the 25-m Dwingeloo telescope (the Netherlands). The latter two telescopes are operated primarily by amateur astronomers, on a volunteer basis. In addition, we complement the HyperFlash data by including lower-cadence observations with the much more sensitive Nan\c{c}ay Radio Telescope (NRT; France) -- which is approximately as sensitive as a $95$-m diameter radio dish. These data were acquired as part of the Extragalactic Coherent Light from Astrophysical Transients (\'ECLAT) observing programme (PI: D.~M.~Hewitt), which targets repeating FRBs. Since the start of 2022, \'ECLAT has been monitoring a sample of roughly 20 repeating FRB sources for about 1\,hr\,week$^{-1}$, each.

\subsubsection{Westerbork RT-1} \label{sec:wb_obs}

At Westerbork, we use an FRB search pipeline that has been described most recently in \cite{ouldboukattine_2025_mnras}. We capture and store raw voltage data (`waveform data' or `baseband data') in the \texttt{VDIF} format with dual circular polarization channels and 2-bit sampling \citep{whitney_2010_ivs}. Subsequently, we convert the baseband data to 8-bit total intensity (Stokes~I) \texttt{SIGPROC} filterbank files using \texttt{digifil} \citep{vanstraten_2011_pasa}. To mitigate dispersive smearing within frequency channels, we created filterbank files at specific time and frequency resolutions that depend on the observed radio frequency range. For observations conducted at P-band (centred at $324$\,MHz), we created filterbank files with a time resolution of $128$\,$\upmu$s and a frequency resolution of $7.8125$\,kHz. At L-band ($1.3$\,GHz central frequency) we used $64$\,$\upmu$s and $15.625$\,kHz resolutions, respectively. When searching the data for bursts, we excise radio frequency interference (RFI) by applying a static mask to remove known affected frequency ranges.
We use \texttt{Heimdall} \citep{Barsdell_phd} to identify burst candidates, where we set a signal-to-noise (S/N) threshold of $7$ and a maximum boxcar width of $1024$ (or $65$\,ms). To minimize the number of false positives, we only search the data within a DM range of $\pm~50$\,\dmunit around the known DM of \rmkt. Initially, we searched around a DM of $480.7$\,\dmunit (which was reported at the time of discovery \citet{tian_2024_atel}), and then later changed this to $464.87$\,\dmunit after we optimized the DM on our initial detection, as described in \citet{ouldboukattine_2024_atel}. Burst candidates were automatically evaluated using the machine learning burst classifier \texttt{FETCH} \citep{agarwal_2020_mnras}. We specifically use models A and H, as previous testing showed they are the most reliable and consistent FETCH models \citep{snelders_2022}, and require that at least one of the two models classifies the candidates with greater $50\%$ probability of being astrophysical in origin. Candidates are manually evaluated if either model exceeds this threshold. As an additional check, regardless of FETCH assigned probability, we also manually inspected all burst candidates that have a DM of $\pm~5$\,\dmunit around the expected DM.
In total, Westerbork observed for $55.1$\,h at P-band and $261.3$\,h at L-band, with a fluence detection limit of $46.5$\,Jy\,ms and $6.6$\,Jy\,ms at P- and L-band, respectively. These limits correspond to a $7\sigma$ detection threshold. Table~\ref{tab:obs_coverage} summarises the observational setup. 

\subsubsection{Stockert}

The Stockert observing and searching strategy has most recently been described in \citet{ouldboukattine_2025_mnras}. At Stockert, we record 32-bit total intensity data using the Pulsar Fast Fourier Transform (\texttt{PFFTS}) backend \citep{barr_2013_mnras}. We then convert these data into 32-bit float filterbanks using the \texttt{filterbank} tool from the \texttt{Sigproc} package. The time and frequency resolutions of the filterbanks are $218.45$\,$\upmu$s and $586$\,kHz, respectively. We search for bursts using tools from the \presto package \cite{ransom_2011_ascl}. We dedisperse the data at $\mathrm{DM}_\mathrm{FRB} = 464.87\,\pm\,10\,$\,\dmunit using \texttt{prepsubband} and mitigate RFI by using \texttt{rfifind}. Using \texttt{single\_pulse\_search.py}, we search for burst candidates with a S/N threshold of $8$ and consider boxcar sizes up to $300$, corresponding to boxcar widths of $\sim65$\,ms. Finally, using \texttt{FETCH}, we then classify the reported burst candidates using models~A and H and a threshold of $>50\,\%$ that the candidate is astrophysical in origin. Stockert observed a total of $142.5$\,h and has a fluence detection threshold of $6.4$\,Jy\,ms (Table~\ref{tab:obs_coverage}). 

\subsubsection{Dwingeloo}

With Dwingeloo, we simultaneously record data at both P-band ($400$--$420$\,MHz) and L-band ($1200$--$1400$\,MHz). At P-band, we record 32-bit filterbank data in two polarization channels (horizontal and vertical) with a time resolution of $192$\,$\upmu$s and a frequency resolution of $0.125$\,MHz across $160$\,channels, providing a total bandwidth of $20$\,MHz. At L-band, we record one channel of $32$-bit filterbank (left-hand circularly polarized) data with a time resolution of $198.4$\,$\upmu$s and a frequency resolution of $0.3125$\,MHz across $640$\,channels, for a total bandwidth of $200$\,MHz. Simultaneously, baseband data is recorded in a $10$~minute buffer, but not searched. These baseband data can be stored if a bright and interesting astrophysical event is detected in the semi-real-time filterbank search. Both the filterbank and baseband (SigMF) data are written using the \texttt{vrt-iq-tools}\footnote{\url{https://github.com/tftelkamp/vrt-iq-tools}} package. The data are searched using tools from the \presto package \citep{Ransom2011PRESTO}. Specifically, \texttt{DDplan.py} is used to create the dedispersion plan for our filterbank files, searching within $\pm 10\%$ of the expected DM, DM$_{\text{frb}} = 480.7\ \mathrm{pc\ cm}^{-3}$ \citep{tian_2024_atel}. The DM$_{\text{frb}}$ was only updated to the more accurately DM measurement, as described in Section~\ref{sec:wb_obs}, after the conclusion of our observational campaign. An \texttt{rfifind} mask is applied for RFI mitigation as \texttt{prepsubband} incoherently dedisperses the filterbank data. The \texttt{single\_pulse\_search.py} script is then used to detect pulses above a S/N threshold of $6$. To reduce the number of candidates by order $\sim$ DM trials, we use DBSCAN clustering to cluster the candidates \citep{db_scan_1996}. If, after clustering, there are more than $800$ candidates, the file is skipped but not deleted. A high number of candidates indicates a large amount of RFI, and the data cannot be searched in real time in such cases. However, the data are retained in case another telescope makes a detection for a potential co-detection. 
Candidates are classified using \texttt{FETCH} model~A \citep{agarwal_2020_mnras}, with a detection threshold of $> 50\%$ probability that the candidate is astrophysical in origin. Candidates that are judged as being promising by \texttt{FETCH} are manually inspected. The full pipeline is open-source and available online\footnote{\url{https://gitlab.camras.nl/dijkema/frbscripts/-/tree/main}}.
Dwingeloo observed for $20.1$\,h at P-band and $17.3$\,h at L-band. The sensitivity of Dwingeloo at P-band is comparable to P-band at Westerbork with a SEFD of $\sim$$2500$\,Jy. At L-band the SEFD is $\sim$$850$\,Jy which is less sensitive compared to Westerbork and Stockert (See Table~\ref{tab:obs_coverage}). The variable and often intense RFI environment, due to the close proximity of the telescope to the ASTRON and JIVE headquarters, can affect the effective detection threshold, leading to fluctuations in burst detectability.

\subsection{\nancay Radio Telescope}

The recording and custom search pipeline employed at the NRT has been described most recently in \cite{konijn_2024_mnras}. At $1.4$\,GHz, the NRT has a system temperature of T$_\text{sys}\approx35\,$K and a gain of $\mathrm{G}\approx1.4\,$K\,Jy$^{-1}$. The low-frequency receiver ($1.1$--$1.8$\,GHz) of the focal plane and receiver system, {\it Foyer Optimis\'e pour le Radio T\'elescope}, was used to conduct the observations. The Nan\c{c}ay Ultimate Pulsar Processing Instrument \citep[NUPPI;][]{desvignes_2011_aipc} recorded 32-bit data with full-polarization information (in a linear basis), $16$\,$\upmu$s time resolution, and $4$\,MHz frequency resolution. The $512$\,MHz observing bandwidth was centred at $1484$\,MHz, and split up into eight subbands, each divided into $16$~channels. To calibrate the polarimetric data, each observation is accompanied by a $10$\,s observation of a $3.33$\,Hz pulsed noise diode. To compensate for the wide channel widths, we applied coherent dedispersion using a DM of $480.7$\,\dmunit, as specified in the discovery ATel \citep{tian_2024_atel}. We updated the DM in our search pipeline to the best-fit value only after the observational campaign was completed. The applied coherent DM differs by $15.84$\,\dmunit from the best-fit DM reported in \cite{ouldboukattine_2024_atel} and used throughout the analysis of the NRT burst sample. This offset introduces intra-channel temporal smearing ranging from $100$\,$\upmu$s in the highest-frequency channel to $283$\,$\upmu$s in the lowest-frequency channel. The NRT carried out $9$~observations for a total of $7.1$\,h, between 6 July 2024 and 24 October 2024, and has a fluence detection threshold of $0.18$\,Jy\,ms (Table~\ref{tab:obs_coverage}).

\section{Analysis and results} \label{sec:ana_res}

\subsection{Burst Properties}

\subsubsection{HyperFlash} \label{sec:hf-method}

To measure the burst properties we used \texttt{filterbank} files. Additionally, to correct for digitisation effects caused by the limited dynamic range of the 2-bit sampling in observations conducted with Westerbork, we apply the scattered power correction (SPC) algorithm to these bursts \citep{vanstraten_2013_apjs}. This technique has been demonstrated in earlier work as described in \cite{ouldboukattine_2025_mnras} and \cite{kirsten_2024_natas}. Bursts detected with Stockert show no digitisation effects owing to the 32-bit recording, whereas the detection with Dwingeloo does not allow accurate burst property measurements due to strong RFI.

\begin{figure*}
    \centering
    \includegraphics[width=\textwidth]{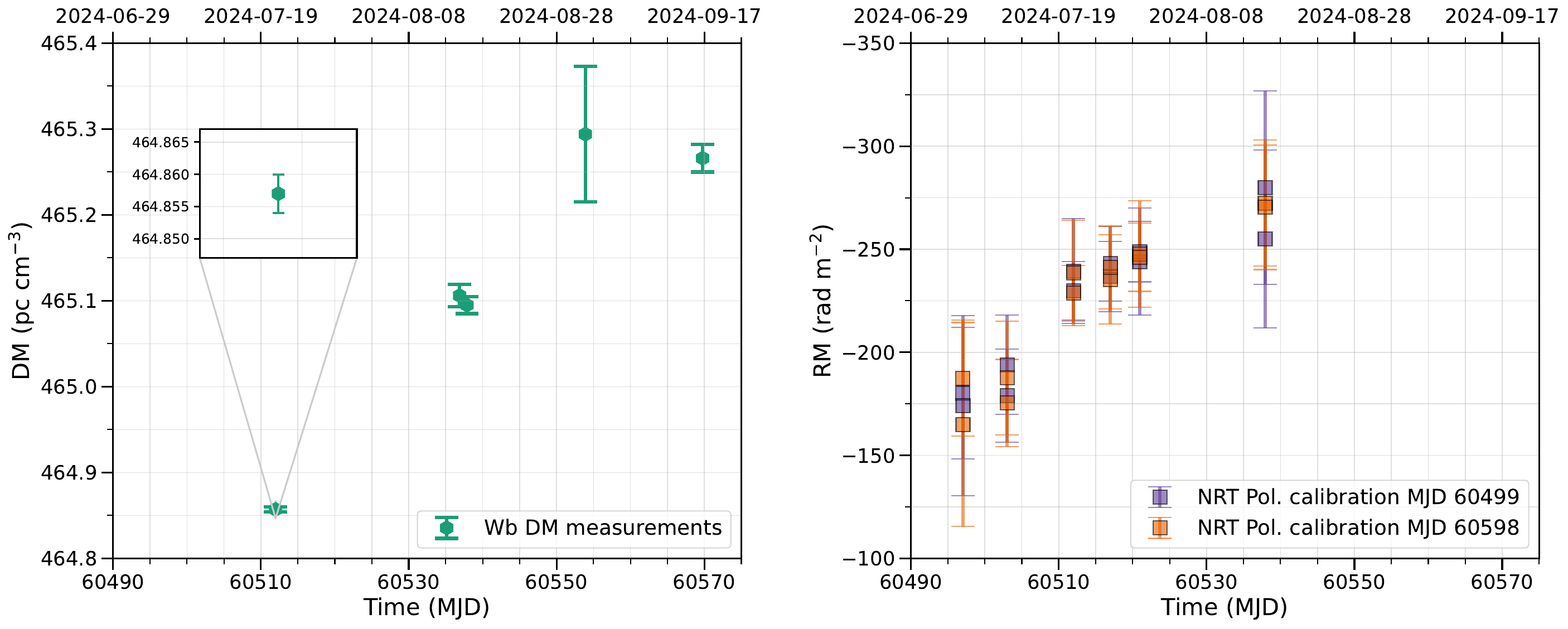}
    \caption{Left: Evolution of the DM for five bursts observed with Westerbork, plotted over time. The baseband data and presence of microstructure enabled precise DM measurements; see Figure~\ref{fig:dyn_matrix} for the corresponding dynamic spectra and S/N versus DM fitting curves. The DM varies by up to $0.41$\,\dmunit over a two-month span. Right: RM of \'ECLAT-detected bursts as a function of time. For each day with detections, the RM was measured for two bursts, to check for consistency. We calibrated the data using two different polarization calibration modelling (PCM) files taken on different dates. The four RM values per day are consistent with each other. We find that the absolute RM increases by $\sim$$80$\,\rmunit over a period of 40 days.}
    \label{fig:dm_rm_over_time}
\end{figure*}

\begin{figure*}
    \centering
    \includegraphics[width=0.95\textwidth]{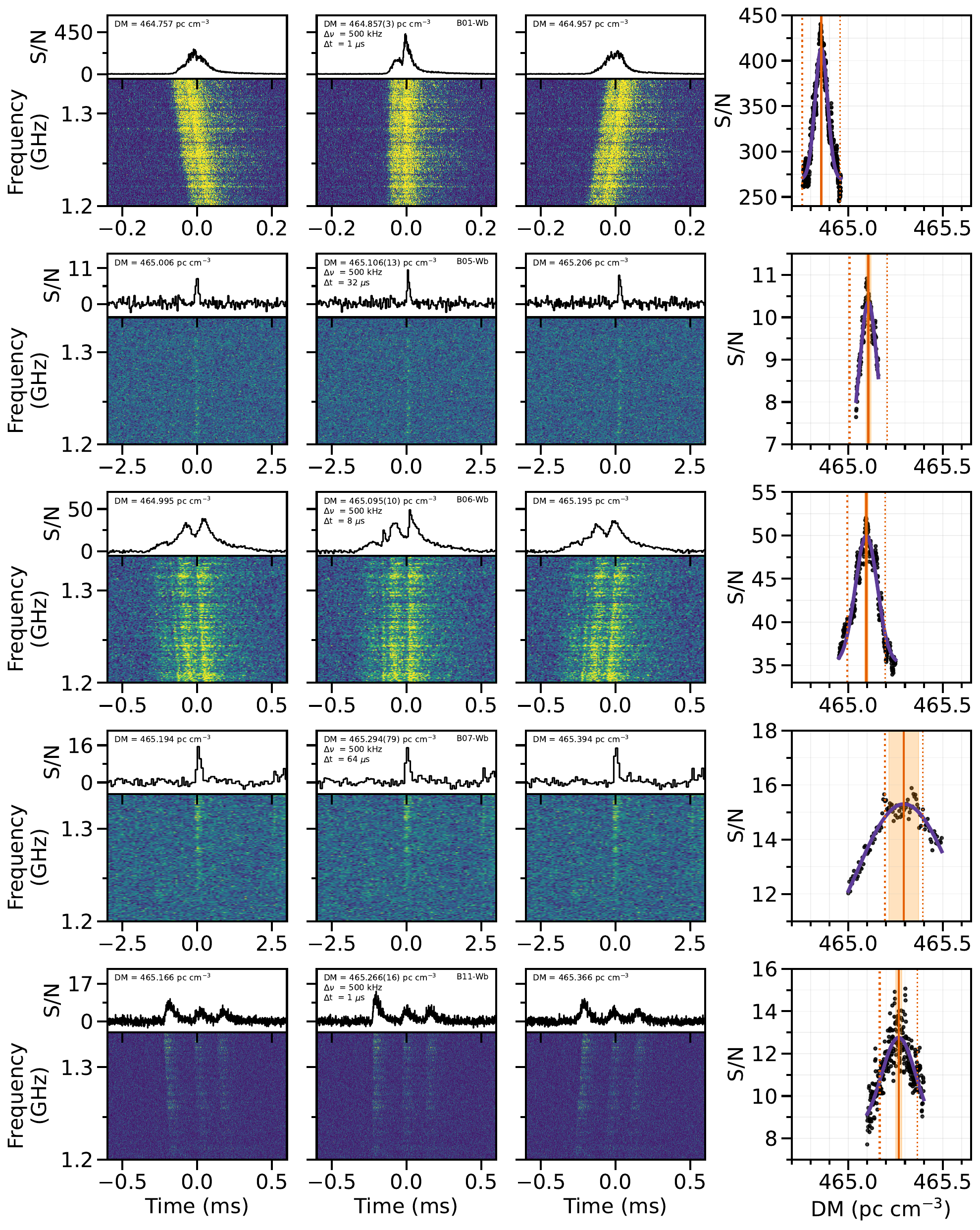}
    \caption{The dynamic spectra and time series for five bursts detected by Westerbork that show fine ($\lesssim 100\,\upmu \textrm{s}$) temporal structure. A single DM value does not appropriately correct for the dispersive delay. All bursts have been coherently dedispersed to the indicated DM and we zoom in on the parts of the bursts that show fine temporal structure. The second column displays the burst spectra and time series using the S/N-optimized DM. The time and frequency resolution used for each filterbank is noted in the top panels. For illustrative purposes, The first and third columns show the bursts dedispersed to the best DM minus and plus $0.1$\,\dmunit, respectively. Finally, the fourth column shows the Gaussian fits to the S/N versus DM curves. The solid orange lines and the light orange regions indicate the best-fit DM and its uncertainty range. The dotted orange lines mark the $\pm\,0.1$\,\dmunit values used in the first and third columns.}
    \label{fig:dyn_matrix}
\end{figure*}

\begin{figure}
    \centering
    \includegraphics[width=1\linewidth]{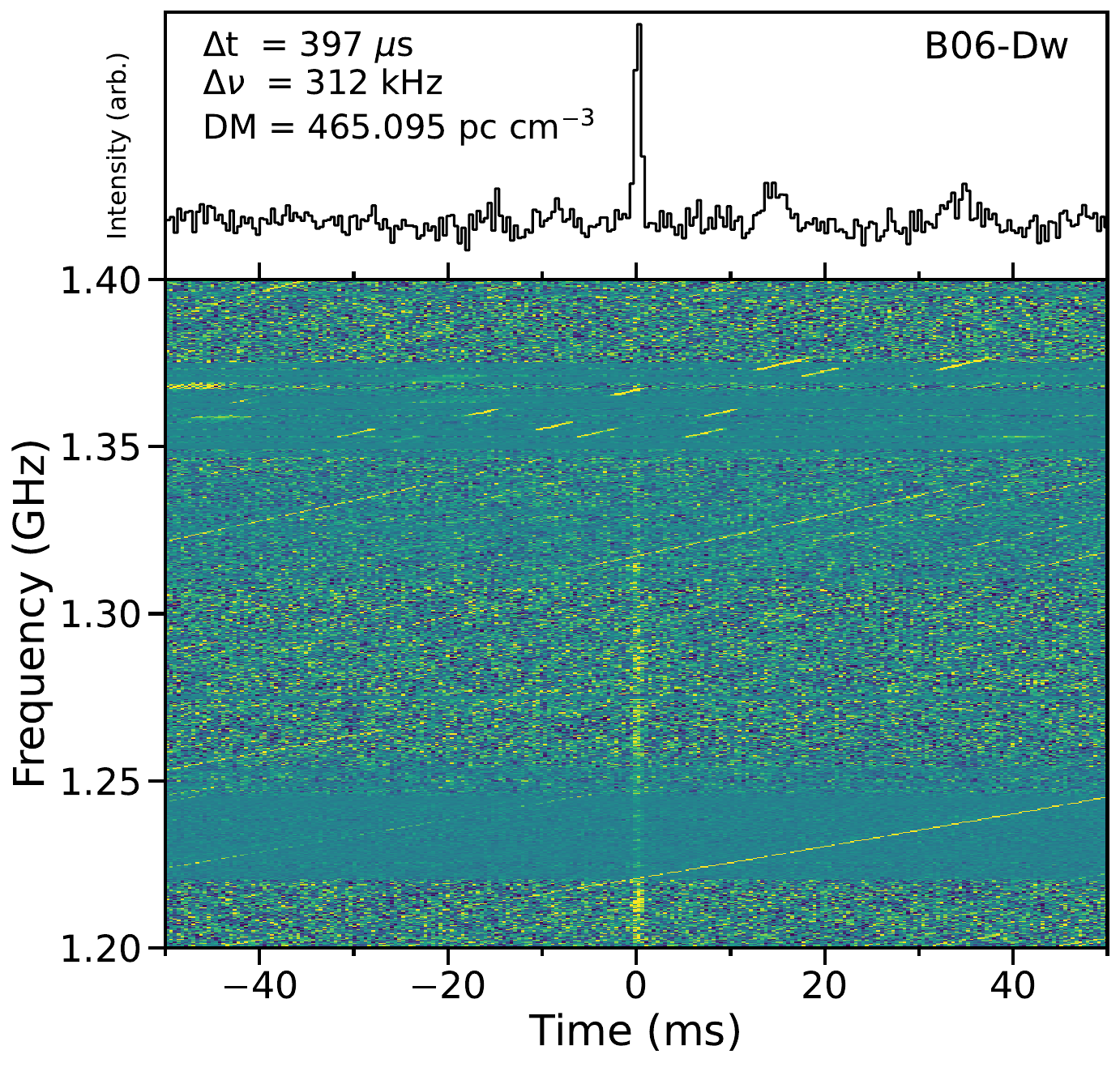}
    \caption{Burst (B06-dw) detected by Dwingeloo. Due to strong RFI in the data it was not possible to measure any burst properties accurately. The DM used is the same as for burst B06-wb as detected by Westerbork, shown in Figure~\ref{fig:dyn_matrix}. This detection of this bright FRB demonstrates the observational capabilities of the Dwingeloo telescope.}
    \label{fig:dwi_burst_b06}
\end{figure}

We correct for the dispersive delay using an optimal DM determined in cases where the bursts showed fine temporal structure ($\lesssim 100$\, $\upmu$s,) also known as `microshots', as listed in Appendix Table~\ref{tab:burst-properties-hf} and shown in the left panel of Figure~\ref{fig:dm_rm_over_time} and Figure~\ref{fig:dyn_matrix}. For bursts where a precise ($\Delta\textrm{DM}<0.1$\,\dmunit), individual DM measurement was not possible, we adopt the value of $464.857$\,\dmunit, corresponding to the DM of the brightest burst, B01-Wb, in our sample. RFI-affected frequency channels are mitigated by manually flagging the channels using tools from the \texttt{PSRCHIVE} software package, such as \texttt{psrzap} and \texttt{pazi}. Additionally, we flag the edges of subbands due to a drop in sensitivity at these frequencies. We manually determine the start and stop times of each burst. We determine the frequency extent of each burst by computing the two-dimensional auto-correlation function (ACF) and fitting a Gaussian function to the frequency axis of the ACF. If twice the full width at half maximum (FWHM) of the Gaussian is greater than or equal to $75\%$ of the total observing bandwidth, we define the burst frequency extent to be equal to the full bandwidth. This condition was met for all bursts in the HyperFlash burst sample. We determine the fluence by summing over the frequency extent and converting the time bins to flux densities using the radiometer equation. Here we assumed a constant system equivalent flux density (SEFD) for all observations, as shown in Table~\ref{tab:obs_coverage}. Finally, we determine the burst time of arrival (ToA) by fitting a Gaussian to the time profile and defining the ToA as the centre of the Gaussian. We report the ToA values as barycentric arrival times in the TDB timescale assuming a dispersion constant of $\mathcal{D} =1/(2.41 \times 10^{-4})$\,MHz$^{2}$\,pc$^{-1}$\,cm$^{3}$\,s with respect to infinite frequency for the measured DM of the burst. For burst B06-Dw, detected with the Dwingeloo telescope (Figure~\ref{fig:dwi_burst_b06}), we were only able to measure the ToA of burst, as strong RFI during the event prevents further analysis.

HyperFlash detected $11$~bursts of which $3$~bursts were detected by more than $1$ telescope. All burst properties are listed in Appendix Table~\ref{tab:burst-properties-hf}. The table is also available in \texttt{.csv} format as part of the supplementary material. An overview of Westerbork-detected bursts for which baseband data are available is shown in Appendix Figure~\ref{fig:wb_bb_family}. All bursts are plotted at the same time and frequency resolution, but on different timescales, highlighting the diversity in burst morphology and the presence of microstructure. Some bursts, such as B02-Wb and B10-Wb, are more than $10$\,ms wide, making it difficult to constrain the DM with high precision. In contrast, other bursts exhibit unresolved emission on microsecond timescales, allowing for precise DM measurements, as discussed further in Section~\ref{sec:dm_var_long}.

\subsubsection{\'ECLAT} \label{sec:nrt-method}

\begin{figure*}
    \centering
    \includegraphics[width=\linewidth]{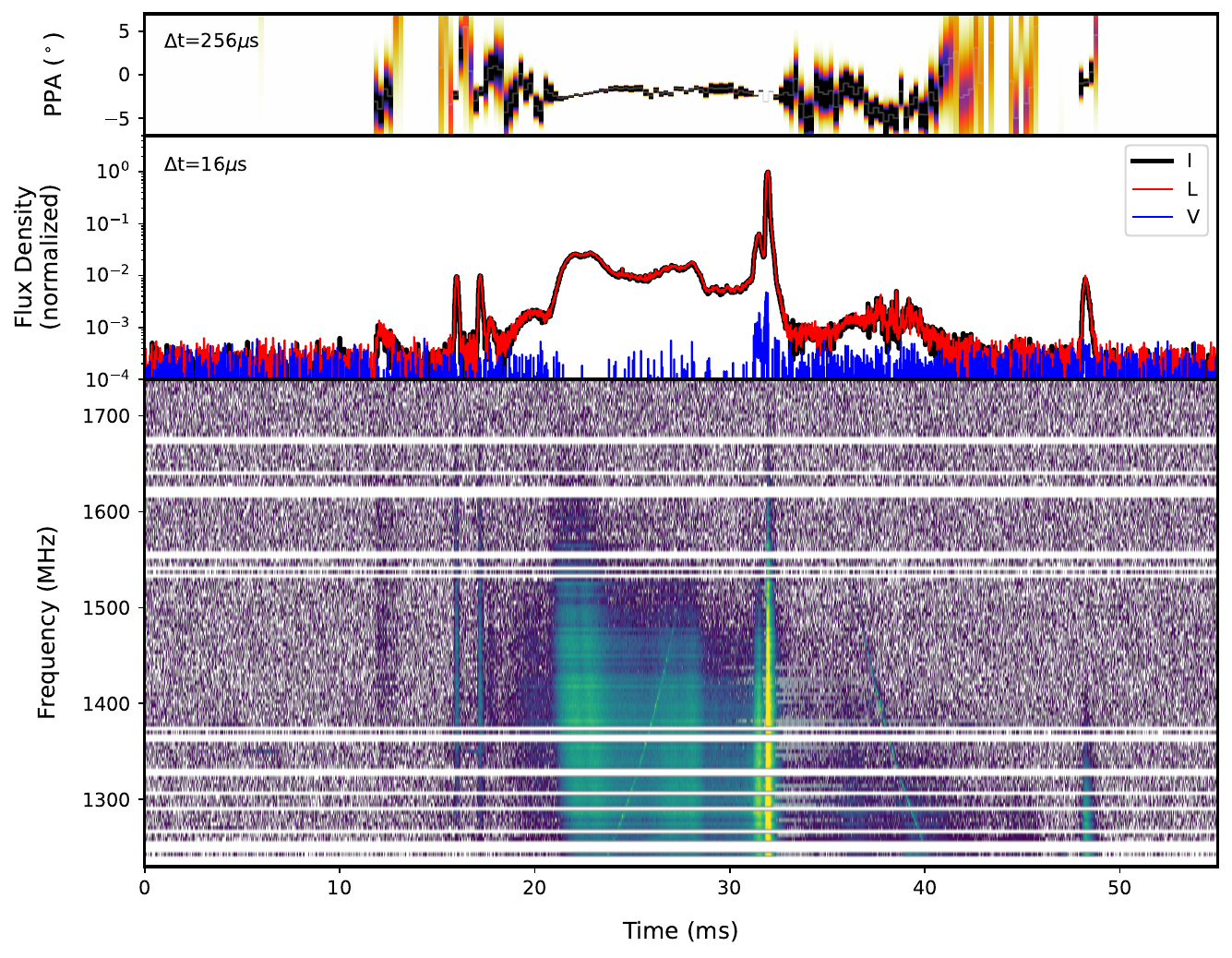}
    \caption{The bottom panel shows the dynamic spectrum of B26-NRT. The colour map is scaled logarithmically. This burst has been coherently dedispersed to a DM of $480$\,\dmunit, and incoherently dedispersed to a DM of $464.86$\,\dmunit (see Section~\ref{sec:nrt-method}). The horizontal white lines indicate channels that have been excised due to the presence of RFI. The middle panel shows the frequency-integrated burst profile in black, also on a logarithmic scale. The linear and circular polarization are shown in red and blue, respectively. The top panel shows the probability density function (for each time bin) of the PPA, which varies by less than a few degrees across the burst duration. The time resolution here has been downsampled by a factor of 16, to account for the DM smearing ($283$\,$\upmu$s in the lowest-frequency channel) due to the difference between the DM that was used for coherent dedispersion and the true DM of the burst.}
    \label{fig:b01_nrt_pol}
\end{figure*}

We dedispersed all the \'ECLAT-detected bursts using a single DM of $464.86$\,\dmunit, determined to be the optimal DM for our brightest burst (B26-NRT) which is shown in Figure~\ref{fig:b01_nrt_pol}. We determined the burst properties using a set of interactive custom Python scripts which were specifically designed for analysis of bursts detected with the NRT. These scripts are publicly available via \texttt{GitHub}\footnote{\url{https://github.com/astroflash-frb/burst_analyzer}}. The first step is to correct the bandpass, by subtracting the mean and dividing by the standard deviation of the off-burst noise on a per-channel basis. Initial RFI flagging was done using \texttt{jess}\footnote{\url{https://github.com/josephwkania/jess}}, where we flagged frequency channels with skewness outliers exceeding $3\sigma$, after which we manually masked any remaining channels contaminated by RFI. The event duration of an individual burst and spectral extent are manually selected. This is done because many bursts have multiple sub-components and/or are not well fit by a simple Gaussian model. The fluence is calculated by summing over the entire observing bandwidth, normalizing using off-burst noise to convert to S/N units, multiplying by the radiometer equation to convert to flux densities, and then integrating over the burst duration. 

We used the Pulsar Archive Calibration program, \texttt{pac}, from the \psrchive tools \citep{hotan_2004_pasa} to perform polarimetric calibration. The expected RM of $-185$\,rad\,m$^{-2}$ \citep{tian_2025_mnras} results in the bursts exhibiting $\lesssim2$ cycles of Faraday rotation across the observing band in Stokes~Q/U, which limits the precision of our RM measurements. Therefore, for each of the six observations with detections, we select the two brightest and most broadband bursts. We used \texttt{rmfit} from \psrchive to search for the RM that maximizes the linearly polarized flux in a range of $-1000$ to $1000$\,rad\,m$^{-2}$. The resulting Faraday Dispersion Functions (FDFs) of the $12$ different bursts can be seen in Appendix Figure~\ref{fig:JoyDivision}. We find that the absolute value of the RM appears to be increasing over time. We then fit a Gaussian function to the main peaks of these FDFs, and quote the centre of the Gaussian as the derived RM value, with the FWHM of Gaussian divided by the S/N as the uncertainty on the RM. These values are also tabulated in Appendix Table~\ref{tab:burst-properties-nrt}.

\begin{figure*}
    \centering
    \includegraphics[width=\linewidth]{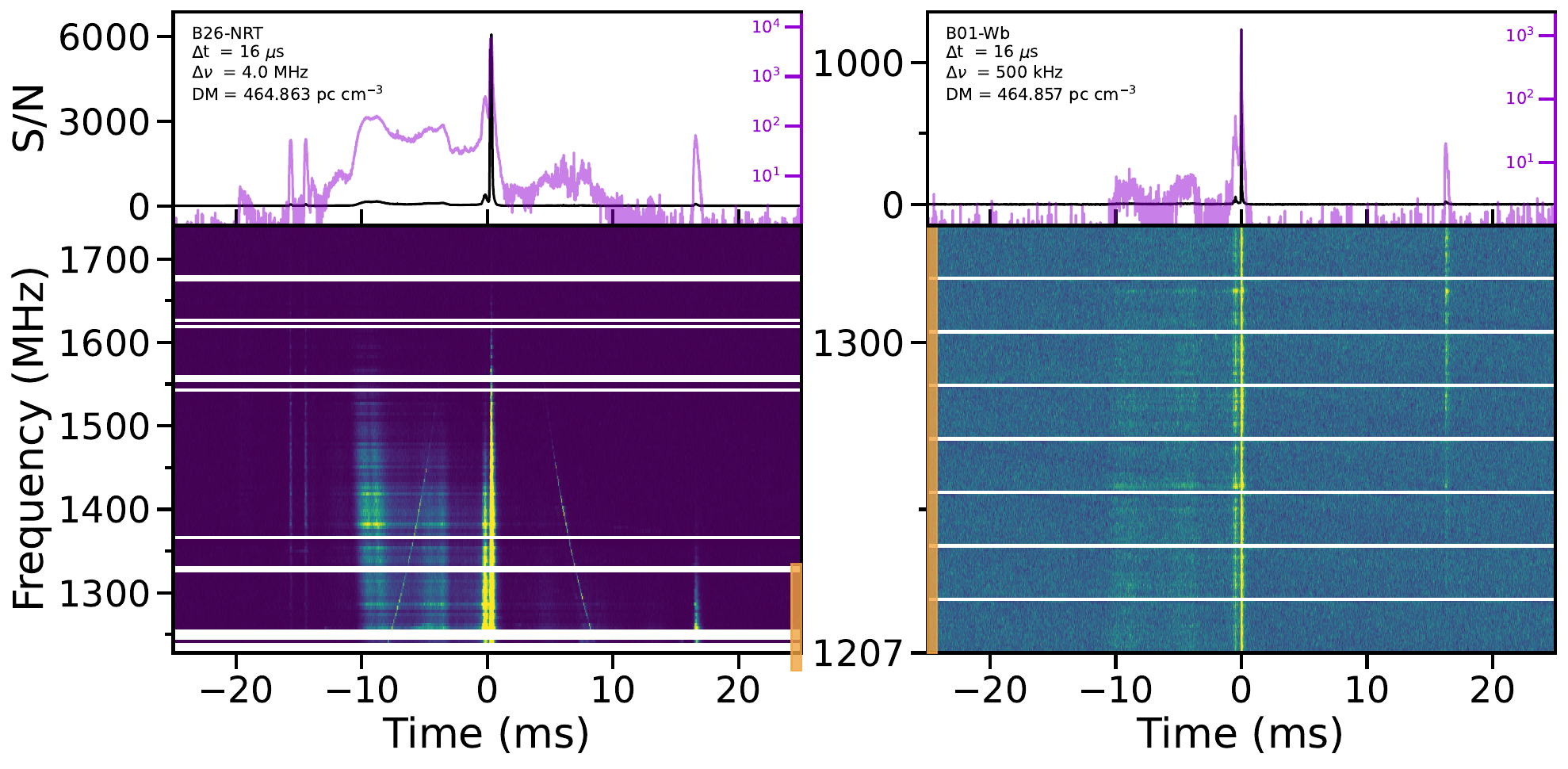}
    \caption{Time series (top panels) and dynamic spectra (bottom panels) for the brightest burst simultaneously detected by \nancay (B26-NRT) and Westerbork (B01-Wb). The orange patches indicate the $128$\,MHz frequency range partially overlapping between the observing setups. The top-left panels show the plotted time and frequency resolutions as well as the applied DM. B26-NRT data was only incoherently dedispersed (between channels), whereas B01-Wb data was also coherently dedispersed (within channels). Although the applied DM values differ slightly between bursts, they remain consistent within the measurement uncertainties (Figures \ref{fig:dyn_matrix} and Appendix~\ref{fig:nrt_dm_opt}). White vertical lines indicate channels zapped due to RFI or are channels at the subband edges.}
    \label{fig:wb_nrt_bright_burst}
\end{figure*}

Figure~\ref{fig:b01_nrt_pol} shows the calibrated polarimetric profile of the exceptionally bright burst B26-NRT, which corresponds to the same event as B01-Wb. We additionally show both bursts side by side in Figure~\ref{fig:wb_nrt_bright_burst} on the same timescale. The linear polarisation has been unbiased and the PDF of the PPA was calculated following the formalism in \citet{everett_2001_apj}. The burst is nearly 100\% linearly polarized, and the PPA varies by a few degrees at most across the burst duration. Upon closer inspection, the PPA exhibits slight temporal variations ($\lesssim1\degree$) around the time of the brightest feature in the burst. There are various factors that we can not perfectly account for that can contribute to subtle variations in the PPA. Firstly, the temporal smearing due to the difference in DM used for incoherent and coherent dedispersion is up to 283\,$\upmu$s (at the lowest-observed frequencies) which is equivalent to the timescale of the tentative PPA variations at the time of the brightest feature. The brightness of this feature also resulted in spectral leakage between channels (`spectral ghosts' are clearly visible around the bright feature). Finally, RFI is likely not perfectly excised due to the broad channels. We thus refrain from interpreting these subtle variations of less than a few degrees in the PPA.

In total, NRT detected $206$~bursts. All burst properties can be found in Appendix Table~\ref{tab:burst-properties-nrt}. Additionally, this table is also available in \texttt{.csv} format as part of the supplementary material.

\begin{table*} 
\caption{\label{tab:obs_coverage}\textbf{Observational set-up}}
\resizebox{\textwidth}{!}{%
\begin{tabular}{c c c c c c S[table-format=2.1] S[table-format=3.1] S[table-format=3.1] S[table-format=3.2,input-decimal-markers={.,}]}
\hline
\hline
{Station$\mathrm{^{a}}$}  & {Band} & {Frequency} & {Bandwidth$\mathrm{^{b}}$} & {Bandwidth per} & {SEFD} & {Detection$\mathrm{^{d}}$} & {Completeness$\mathrm{^{e}}$} & {Time observed$\mathrm{^{f}}$} & {Time observed$\mathrm{^{g}}$} \\
 &  & {[MHz]} & {[MHz]} & {subband [MHz]} & {[Jy]} & {threshold [Jy~ms]} & {threshold [Jy~ms]} & {[hr]} & {[hr]} \\
\hline
Wb  & P$_{\rm Wb}$    & 300--356          &50     & 8   & 2100$\mathrm{^{c}}$  & 46.5 & 172.5 & \multicolumn{1}{c}{--} & 55.1 \\
Dw  & P$_{\rm DW}$    & 400--420          &20     & 20  & $\sim 2100$  & \multicolumn{1}{c}{--}  & \multicolumn{1}{c}{--} & \multicolumn{1}{c}{--}    & 20.1 \\
Dw  & L$_{\rm Dw}$    & 1200--1400        &180    & 100 & $\sim 850$   &  \multicolumn{1}{c}{--} &  \multicolumn{1}{c}{--} & \multicolumn{1}{c}{--}   & 17.3 \\
Wb  & L$_{\rm Wb}$    & 1207--1335        &100    & 16  & 420$\mathrm{^{c}}$   & 6.6  & 24.4  & 166.1   & 261.3 \\
NRT & L$_{\rm NRT}$   & 1228--1740        &500    & 512 & 25    & 0.18  & 0.65 & 5.5 & 7.8 \\
St  & L$_{\rm St}$    & 1332.5--1430.5    &90     & 98  & 385   & 6.4  &  23.6 & 117.6 & 142.5 \\
\hline
\multicolumn{9}{l}{Total time at $1.4\,\mathrm{GHz}$ (L-band) on source [hr]$\mathrm{^{g}}$} & \textrm{323.9} \\
\hline
\multicolumn{9}{l}{Total telescope time/total time on source [hr]$\mathrm{^{h}}$} & \textrm{504.1/353.9} \\
\hline

\multicolumn{9}{l}{$\mathrm{^{a}}$ Wb: Westerbork RT1, St: Stockert, Dw: Dwingeloo, NRT: \nancay radio telescope} \\
\multicolumn{9}{l}{$\mathrm{^{b}}$ Effective bandwidth accounting for RFI and band edges.} \\
\multicolumn{9}{l}{$\mathrm{^{c}}$ From the \href{https://www.evlbi.org/sites/default/files/shared/EVNstatus.txt}{EVN status page}.} \\
\multicolumn{9}{l}{$\mathrm{^{d}}$ Assuming a $7\sigma$ detection threshold and a typical FRB pulse width of $1~\mathrm{ms}$.} \\
\multicolumn{9}{l}{$\mathrm{^{e}}$ Assuming a $15\sigma$ detection threshold and a width of $3~\mathrm{ms}$.} \\
\multicolumn{9}{l}{$\mathrm{^{f}}$ On-source hours between MJD $60497$ and $60569$ (6 July – 16 September 2024) used for the cumulative burst energy distribution shown in Figure \ref{fig:cumulative_burstrate}.} \\
\multicolumn{9}{l}{$\mathrm{^{g}}$ Total on-source hours during the full observational campaign between MJD $60496$ and $60602$ (5 July – 19 October 2024).} \\
\multicolumn{9}{l}{$\mathrm{^{h}}$ Total time on source accounts for overlap between the participating telescopes.} \\
\end{tabular}
}
\end{table*}

\newpage

\subsection{DM variation between bursts} \label{sec:dm_var_long}

Quantifying DM variation between bursts provides a means to probe the slight variations in the line of sight or the local environment of a repeating FRB source. We record the raw voltage data for Westerbork observations, which allows us to create data products with a time resolution as high as $31.25$\,ns. We use the Super~FX~Correlator (\sfxc) to coherently dedisperse and channelise our data \citep{keimpema_2015_exa}. Five of the seven bursts detected with Westerbork show structure on tens of microsecond timescales, see Appendix Figure~\ref{fig:wb_bb_family}. To optimize for the DM, we create coherently dedispersed filterbank files for a range of trial DM values, with steps of $0.001-0.01$\,\dmunit depending on the brightness and timescale of the microstructure components. This method prevents residual smearing within a channel. For each trial DM, we then measure the peak S/N for the brightest feature of the burst. We then fit a Gaussian function to the resulting S/N versus DM curve to determine the optimal DM and its uncertainty. We were not able to constrain the DM of two bursts (B02-Wb and B10-Wb) with high accuracy ($\Delta\textrm{DM}<0.1$\dmunit) due to the lack of microshots; therefore, we adopt the DM of the brightest burst (B01-Wb) for them. 

In Figure~\ref{fig:dyn_matrix} we show the resulting S/N versus DM curve for five bursts. Additionally, we show the coherently dedispersed dynamic spectrum for each burst, at their `best' value and $\pm0.1$\,\dmunit. The figure shows that a single DM value does not appropriately correct for the dispersive delay for all five bursts . In the left panel of Figure~\ref{fig:dm_rm_over_time}, we show the measured DM values for each burst as a function of time. We find that DM varies between $464.857 \pm 0.003$ and $465.266 \pm 0.016$\,\dmunit, which is a total change of $0.41 \pm 0.016$\,\dmunit, over a period of approximately two months. 

\subsection{Apparent DM variations within a single burst}

During our observing campaign, the brightest burst in our sample was simultaneously detected with Westerbork and \nancay, B01-Wb and B26-NRT, respectively (Figure~\ref{fig:wb_nrt_bright_burst}) \citep{ouldboukattine_2024_atel}. Using the Westerbork data, we precisely measured the DM to be $464.857 \pm 0.003$\,\dmunit\ (see Section~\ref{sec:dm_var_long}). We also optimize the DM for the NRT-detected burst (Figure~\ref{fig:b01_nrt_pol}). We dedispersed the burst over a range of trial dispersion measures and measured the peak S/N of the brightest component at each step. We then fit a Gaussian function to the resulting S/N curve to determine the optimal DM. The fit, shown in the left column of Appendix~Figure~\ref{fig:nrt_dm_opt}, yields a best-fit DM of $464.863 \pm 0.029$\,\dmunit. This value is consistent with the aforementioned DM derived from the Westerbork data, as shown in the top row of Figure~\ref{fig:dyn_matrix}.

After correcting for dispersion, we find that the burst has a component that shows an apparent residual drift, starting at $\sim$$20$\,ms in the dynamic spectrum of Figure~\ref{fig:b01_nrt_pol}. The other shorter components of this burst, occurring before and after this drifting feature, do not show any evidence for such a drift. The drifting component shows an intensity dip that is most visible in the time series. To quantify apparent drift, we optimized the DM by fitting the narrow intensity dip (right column of Appendix~Figure~\ref{fig:nrt_dm_opt}) with an upside down Gaussian. We find the component straightens out at a DM of $465.540 \pm 0.028$\,\dmunit. This DM differs by $+0.677$\,\dmunit compared to the DM that straightens all other sub-bursts.

\subsection{Scattering analysis}\label{sec:scattering}

Three bursts in the sample show frequency-dependent asymmetries suggestive of pulse broadening due to multipath propagation. We infer the scattering delays $\tau$ of these bursts by modelling the burst profiles as the convolution a Gaussian pulse (or series of Gaussian pulses) with a one-sided exponential that has a $1/e$ delay $\tau$ that scales with observing frequency as $\tau~\propto~\nu^{-4}$ (as strictly required for the assumption of a Gaussian scattered image; \citealt{lambert99,geiger25}). All of the bursts we consider have multiple components, which we model as the sum of $N$ Gaussians each convolved with a one-sided exponential. The number of sub-components $N$ included in the model is kept to the smallest number possible while minimizing $\chi^2$. We also fix $\tau$ to be constant across all sub-components in an individual burst. To fit for $\tau$, we evaluate the mean burst profile in multiple frequency subbands, where the number of subbands is set to yield a peak $\rm~S/N~>~10$. To reduce the number of free parameters, the mean arrival times of each sub-component are fixed across all subbands. The Gaussian amplitudes of each sub-component are allowed to vary freely between subbands, while the Gaussian widths are only allowed to vary slightly (by $<20\%$). 

\begin{figure*}
    \centering
    \includegraphics[width=\textwidth]{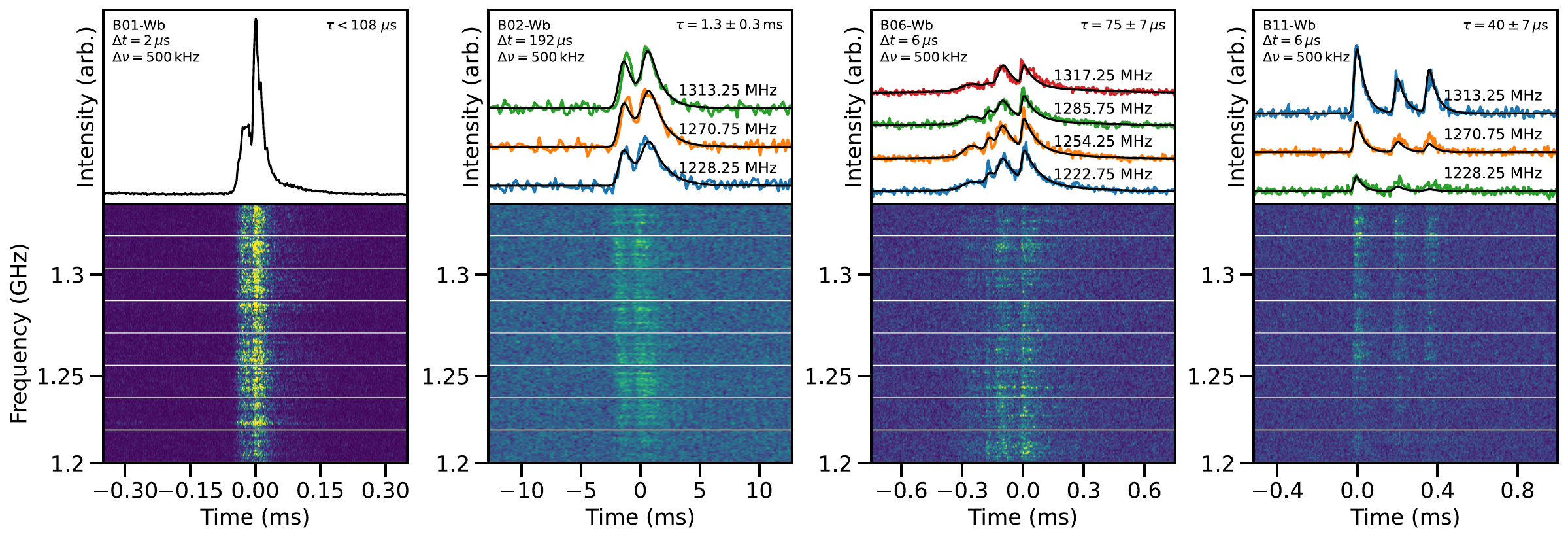}
    \caption{Pulse broadening constraints for four Westerbork bursts. Bottom panels show dynamic spectra and upper panels show frequency-averaged burst profiles. In the upper panels we also indicate the burst ID, as well as the time and frequency resolution of each burst. The three right-hand bursts show evidence of chromatic pulse broadening consistent with scattering delays $\tau$ noted on each panel; each $\tau$ is referenced to the lowest-frequency subband. Black curves show the best-fit burst models (see Section~\ref{sec:scattering}). The first burst, B01-Wb, instead provides a stringent $95\%$ confidence upper limit on the scattering delay based on the empirical width of its ACF, $\tau < 108$\,$\upmu$s at 1271\,MHz.}
    \label{fig:tau}
\end{figure*}

Figure~\ref{fig:tau} shows the best-fit burst profiles for the three bursts with evidence of pulse broadening. We infer scattering delays $\tau_\textrm{B02-Wb} = 1.3 \pm 0.3$\,ms at $1228.25$\,MHz, $\tau_\textrm{B06-Wb} = 75\pm7$\,$\upmu$s at $1222.75$\,MHz, and $\tau_\textrm{B11-Wb} = 40\pm7$\,$\upmu$s at $1313.25$\,MHz for these three bursts, respectively. Scaling $\tau$ to $1$\,GHz assuming $\tau \propto \nu^{-4}$ yields respective scattering delays $\tau_\textrm{B02-Wb}(1\ {\rm GHz}) = 2.9\pm0.7$\,ms, $\tau_\textrm{B06-Wb}(1\ {\rm GHz}) = 170\pm16$\,$\upmu$s, and $\tau_\textrm{B11-Wb}(1\ {\rm GHz}) = 120\pm21$\,$\upmu$s. We note that one of these bursts (B06-Wb) is best fit by a model with five sub-components, and the $1\sigma$ error quoted on $\tau$ for this burst is likely an underestimate. Regardless, these results suggest that the scattering delay may vary significantly between bursts. 

The top-left panel of Figure~\ref{fig:tau} also shows the frequency-averaged burst profile of an extremely bright burst (B01-Wb) that does not show clear evidence of frequency-dependent pulse broadening. Evaluating the temporal ACF of this burst (following methods in \citealt{ocker_2023_mnras}) yields an empirical upper limit on the burst FWHM of $51 \pm 28$\,$\upmu$s, which gives a 95\% confidence upper limit on $\tau$, $\tau_\textrm{B01-Wb} < 108$\,$\upmu$s at 1271\,MHz, equivalent to $\tau_\textrm{B01-Wb} < 324$\,$\upmu$s at $1$\,GHz (assuming $\tau \propto \nu^{-4}$). This empirical upper limit is almost $10$~times smaller than the largest scattering delay we infer, which lends additional evidence of scattering variability. 

\subsection{Scintillation analysis}
\label{sec:scint}
For the scattering measurements described in Section\,\ref{sec:scattering}, we estimate that the corresponding scintillation bandwidth from that same scattering medium is on the order of $0.1$--$1$\,kHz. The Galactic electron density model \texttt{NE2001} \citep{cordes_2002_arxiv} estimates a ISM scintillation bandwidth of $655.7$\,kHz (referenced to $1.271$\,GHz) for this line of sight (see Table\,\ref{tab:rmkt-properties}). Therefore, the $4$\,MHz frequency resolution of the NRT data is too coarse to be sensitive to the expected scintillation scales. As a result, we focus on analysing the Wb RT-1 baseband data, at sufficiently high frequency resolution. We create filterbank data from the baseband data with time and frequency resolutions depending upon the duration and morphology of the burst, as listed in Table~\ref{tab:scint}. Using this data, we compute frequency spectrum ACFs, which we fit for both a single and double Lorentzian function, following the methodology outlined in \citet{nimmo_2025_natur}. We find an approximate $\sim$$100$\,kHz frequency scale consistent across all $7$\,bursts that we analyse, see Table~\ref{tab:scint}. While we do not see evidence for strong frequency evolution of this $\sim$$100$\,kHz scale (unsurprising given that our observations are at relatively high frequencies), the consistency across all bursts leads us to rule out self-noise as the cause of this frequency scale, which would change burst to burst given the drastically different burst morphologies observed. Furthermore, we find no evidence for an equivalent frequency scale in similar analysis of the off-burst data, leading us to conclude that this scale is not an instrumental or analysis artefact. The slight burst-to-burst variations in the measured scintillation scale can be attributed to differences in burst S/N and may also be due to refractive scintillation on week-month timescales comparable to the separation between these bursts \citep[e.g.,][]{daszuta_2013_mnras}. We therefore conclude that the $\sim$$100$\,kHz frequency scale can be attributed to scintillation. 

\begin{figure}
    \centering
    \includegraphics[width=\columnwidth]{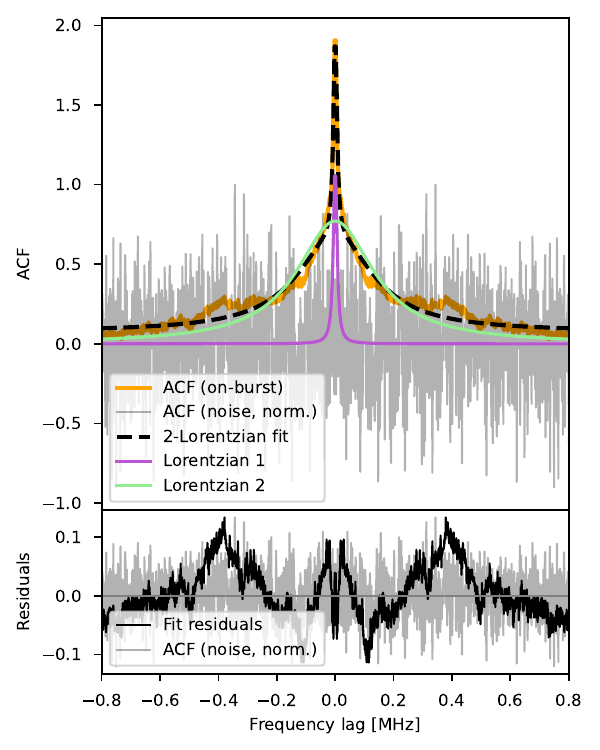}
    \caption{
    The autocorrelation function (ACF) of burst B01-Wb, shown in orange, using the Wb RT-1 baseband data, channelised to a resolution of $60$\,Hz and downsampled by a factor of $16$ to $0.97$\,kHz. In the top panel we show the double Lorentzian fit in black, and we attribute both fitted scales to scintillation as described in Section~\ref{sec:scint}, with the two individual Lorentzian components shown in green and purple. In the bottom panel the black line shows the residual of the double Lorentzian fit. In both panels the off-burst ACF is shown in grey and scaled to have a peak of 1 for easy comparison with the on-burst ACF.}
    \label{fig:scint}
\end{figure}

For the brightest burst in our sample, B01-Wb, we measure a scintillation bandwidth of $159\pm8$\,kHz and a modulation index of $0.6$ (at $1.271$\,GHz, the centre of the observing band). A second, narrower frequency scale is also present in the ACF of this burst (Figure~\ref{fig:scint}). To explore narrower frequency scale further, we analyse bursts B01-Wb, B02-Wb, B06-Wb, B10-Wb, and B11-Wb at even higher frequency resolution, given that the burst structure and baseband data allows us to probe sub-kHz scales. We create ACFs of burst spectra at a resolution of 120\,Hz (60\,Hz for B01-Wb), which is roughly the inverse of the temporal width of each burst. We find a $\sim$kHz scale in all bursts. However, structure on the same frequency scale is evident in the off-burst data, adding doubt to the astrophysical nature of this frequency scale. B01-Wb, however, shows an additional $6.4\pm0.9$\,kHz scale, with a modulation index of $1.02\pm0.1$ at 1.271\,GHz. For this scale we find a hint of a shallower frequency dependence, though consistent with the $\nu^{\alpha=-4}$ evolution expected for scintillation: $\alpha=2.3\pm1.4$, by dividing the burst spectrum up into $16$ subbands and measuring the frequency scale in each subband independently (see Appendix Figure~\ref{fig:scin_freq_depen}). This is consistent with the upper limit of scatter broadening we measure for B01-Wb, $\tau_\textrm{B01-Wb}<108$\,$\upmu$s, which implies a scintillation bandwidth $\Delta\nu_{\rm scint}\gtrsim1.5$\,kHz. We therefore attribute the $6.4\pm0.9$\,kHz scale to be scintillation from a second screen along the line of sight to \rmkt.

\begin{table}
\caption{\textbf{Scintillation measurements for the Wb RT-1 bursts}}
\label{tab:scint}
\resizebox{\columnwidth}{!}{%
\begin{tabular}{
  c
  S[table-format=4.0]   
  S[table-format=1.2]   
  S[table-format=3.0(2),separate-uncertainty=true] 
  S[table-format=1.2(2),separate-uncertainty=true] 
  S[table-format=1.1(1),separate-uncertainty=true] 
  S[table-format=1.2(2),separate-uncertainty=true] 
}
\toprule
\toprule
{Burst ID} & {$\Delta$Time} & {$\Delta$Freq} & {Galactic $\Delta\nu_\mathrm{scint}$} & {Galactic $m^\mathrm{a}$} & {Extragalactic $\Delta\nu_\mathrm{scint}$} & {Extragalactic $m^\mathrm{a}$} \\
{[Wb]} & [$\upmu$s] & {[kHz]} & {[kHz]} & & {[kHz]} & \\ 
\midrule
B01 & 128  & 3.91          & 159 \pm 8  & 0.62 \pm 0.01 & 6.4 \pm 0.7 & 1.05 \pm 0.02 \\
B02 & 4096 & 1.95$^\mathrm{b}$ & 112 \pm 4  & 1.06 \pm 0.01 & \multicolumn{1}{c}{--} & \multicolumn{1}{c}{--} \\
B05 & 64   & 7.81          & 99 \pm 16  & 1.70 \pm 0.05 & \multicolumn{1}{c}{--} & \multicolumn{1}{c}{--} \\
B06 & 256  & 1.95          & 91 \pm 3   & 0.95 \pm 0.01 & \multicolumn{1}{c}{--} & \multicolumn{1}{c}{--} \\
B07 & 64   & 7.81          & 96 \pm 16  & 1.40 \pm 0.04 & \multicolumn{1}{c}{--} & \multicolumn{1}{c}{--} \\
B10 & 4096 & 1.95$^\mathrm{b}$ & 74 \pm 3   & 1.13 \pm 0.01 & \multicolumn{1}{c}{--} & \multicolumn{1}{c}{--} \\
B11 & 128  & 3.91          & 121 \pm 5  & 1.06 \pm 0.01 & \multicolumn{1}{c}{--} & \multicolumn{1}{c}{--} \\
\bottomrule
\multicolumn{7}{l}{$\mathrm{^{a}}$Modulation index} \\
\multicolumn{7}{l}{$\mathrm{^{b}}$Downsampled from $0.122$\,kHz} \\
\end{tabular}%
}
\end{table}

\subsection{Cumulative burst rates}

The cumulative distribution of burst energies can be fit using a power law, $R (> \mathcal{F}) \propto \mathcal{F}^{\gamma_\textrm{C}}$. Where, $R$ is the rate of bursts, typically expressed in units per hour. $\mathcal{F}$ is the fluence of the bursts and $\gamma_\textrm{C}$ is the slope of the cumulative distribution. In the literature, such as \cite{ouldboukattine_2025_mnras}, the burst energies are expressed in specific burst energy (erg\,Hz$^{-1}$). The specific burst energy takes account of the redshift and therefore distance to the host galaxy of the FRB. It also enables fairer comparisons between different telescopes, which often observe with different bandwidths and central frequencies. In turn, this also allows for direct comparison between different FRB repeaters. At the time of writing, there is not yet a measured redshift for \rmkt. We therefore fit the power law to the cumulative distributions in terms of fluence.

\begin{figure}
    \centering
    \includegraphics[width=\columnwidth]{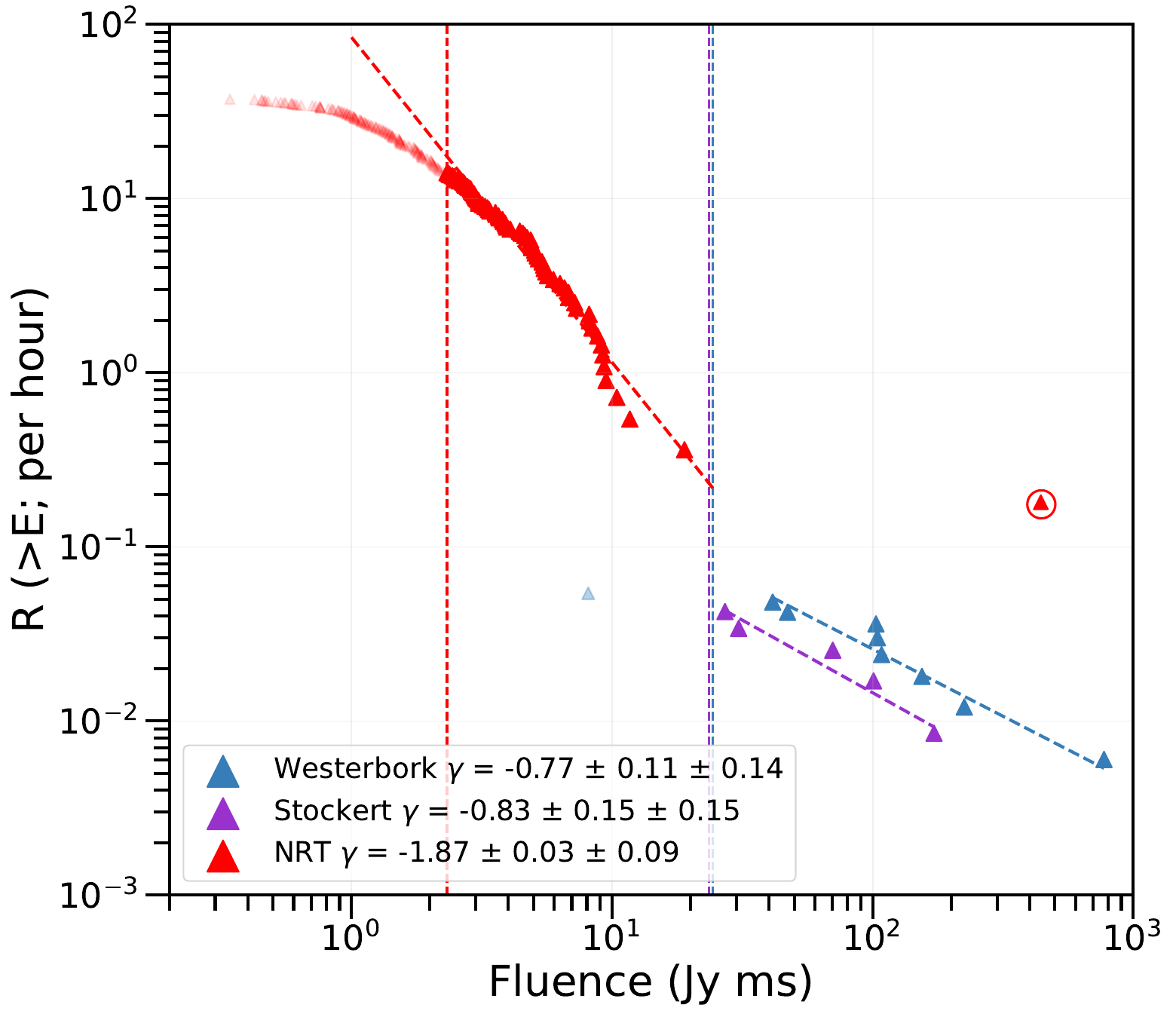}
    \caption{
    Cumulative burst rate distribution for Westerbork, Stockert, and \nancay (NRT) at $1.4$\,GHz (L-band). Comparing the different rates reveals a break in the distribution towards higher energies. The burst rates are calculated based on the activity of the source between MJD~$60497$ (first detection) and MJD~$60570$ (last detection). The red dotted line for NRT marks the turnover point in the distribution where a single power law provides the best fit, as determined using the \texttt{powerlaw} Python package. The transparent data points to the left of the vertical line are excluded from the fit, as is the single circled high-energy point, which is shown only for illustrative purposes to indicate consistency with a break in the distribution. The purple and blue dotted lines represent the completeness thresholds for Stockert and Westerbork, respectively. A $20\%$ uncertainty is assumed on the burst energies. We report two error bars: the first denotes the $1$-$\sigma$ statistical uncertainty, while the second reflects the $1$-$\sigma$ uncertainty derived from a bootstrapping method.}
    \label{fig:cumulative_burstrate}
\end{figure}

We fit power laws to the cumulative burst distributions observed with NRT, Westerbork, and Stockert, as shown in Figure~\ref{fig:cumulative_burstrate}. We restrict the analysis to the period between the first detected burst by NRT on 2024 July 6 (MJD~$60497$) and the last detected burst by Westerbork on 2024 September 16 (MJD~$60569$), in order to enable a consistent comparison of the burst rates across the different observational campaigns. The total exposure time for each telescope during this interval is listed in Table~\ref{tab:obs_coverage}. To fit a power law to the distribution from \nancay, we first determine the turnover point in the distribution above which the data is best described by a single power law. We do this using the Python package \texttt{powerlaw} \citep{alstott_2014_ploso}. We find the turnover point to be at $2.3$\,Jy\,ms. For Westerbork and Stockert we fit a power law for bursts with energies higher than the completeness threshold. The various detection and completeness thresholds are indicated in Table~\ref{tab:obs_coverage}. Next, we make an initial guess of the power-law slope ($\gamma_\textrm{C}$) using a maximum likelihood method, as described in \cite{crawford_1970_apj} and \cite{james_2019_mnras}. Using \texttt{scipy.optimize.curvefit}, we fit a power law and assume a $20\%$ error on the fluence values, which arises from the uncertainty in the SEFD of each telescope. We report two uncertainties in our results: the first is the 1-$\sigma$ error derived by \texttt{curvefit}, and the second is obtained by resampling method to assess the variance of the fit. To estimate the variance of the fit, we repeatedly draw random subsets containing $90\%$ of the bursts without replacement and refit the power law 1000 times.

We observe a break in the cumulative distribution, where the slope of \nancay is steeper than those of Westerbork and Stockert. For the fluences of \nancay, we find a steep power law of $\gamma_\textrm{C}^{\textrm{NRT}} = -1.87 \pm 0.03 \pm 0.09$. We do not attempt to account for the single bright burst detected in the \nancay sample, as there are insufficient data points to fit a broken power law. Nevertheless, we include this data point in Figure~\ref{fig:cumulative_burstrate} for illustrative purposes, to highlight that the detection is consistent with a break in the fluence distribution. For Westerbork and Stockert, we find flatter power laws of $\gamma_\textrm{C}^{\textrm{Wb}} = -0.69 \pm 0.12 \pm 0.15$ and $\gamma_\textrm{C}^{\textrm{St}} = -0.83 \pm 0.15 \pm 0.16$, respectively.

\section{Discussion} \label{sec:discussion}

The high sensitivity, large bandwidth and accurate polarization calibration of NRT allows for the precise determination of burst morphology and RM, while the ability to record complex voltages with Westerbork allows for the removal of intra-channel dispersion and upchannelization to precisely characterize DM and scattering/scintillation. We use the combined information from our bursts detected at the two telescopes to constrain the local environment of \rmkt. We characterize the energy distribution of the source, using all bursts detected at NRT, Westerbork and Stockert.

\newpage

\subsection{Dispersion and rotation measure variability}

By measuring the DM to high precision (order of $0.2$--$0.001$\,\dmunit) for $5$~bursts, using the method explained in Section~\ref{sec:dm_var_long} and shown in Figure~\ref{fig:dyn_matrix}, we measure an increase of the DM of $+~0.41$\,\dmunit over two months. 

Deriving accurate FRB DMs is challenging because, in the absence of microshots, dispersion and drifting effects such as downward drifting sub-bursts \citep[the `sad trombone' effect;][]{hessels_2019_apjl,faber_2024_apj}, are difficult to disentangle. In cases with sufficient signal, one can optimize the temporal structure using a structure-optimizing algorithm such as \texttt{DM-phase} \citep{seymour_2019_ascl}. However, when attempting to find the best DM for faint bursts where downward drifting is not clearly visible, it is possible to overestimate the DM. This can lead to a wide spread in measured DM values, which might be mistakenly interpreted as DM variability. Nonetheless, accurate and precise DMs can be achieved at $1.3$\,GHz when an FRB displays fine temporal structure ($\lesssim 100$\,$\upmu$s) with high S/N.
Pulsars typically show DM variability on the order of $0.001$--$0.01$\,\dmunit over timespans of months to years \citep[e.g.,][]{petroff_2013_mnras}. Scaling these values naively from a $\sim$$1$\,kpc pulsar line-of-sight to a typical $\sim$$10$\,kpc Milky Way crossing suggests a possible Galactic contribution in these timespans of order $\lesssim0.01$--$0.1$\,\dmunit. These DM variations are still smaller compared to the DM variation we observe for \rmkt. We therefore attribute the large DM variations reported here primarily to a turbulent local magneto-ionic environment, while noting that a small residual Galactic contribution at the level estimated above cannot be ruled out.

Long-term monitoring of repeating FRBs enables the study of the DM evolution/stability. This behaviour varies among repeaters. For example, \rthree exhibits a fairly stable DM, with three years of CHIME/FRB data showing stochastic variations of no more than $\leq 0.8$ \dmunit\ \citep{mckinven_2023_apj}. Similarly, for \meightone, \cite{nimmo_2023_mnras} found that the DM remained stable within $< 0.15$\,\dmunit over observations spanning nearly a year. In contrast, for \rone, \cite{li_2021_natur} found that the DM varies and increases at a rate of $+0.85 \pm 0.1$\,\dmunityr, based on a comparison of best-fit DM values from studies spanning more than seven years, between 2013 and 2020 \citep{hessels_2019_apjl,oostrum_2020_aa,jahns_2023_MNRAS}. Following the reported increases, the DM of \rone is now decreasing, with both \cite{wang_2022_atel_15619} ($552.5 \pm 0.9$\,\dmunit\ in September 2022) and \cite{snelders_2025_arxiv} ($551.92 \pm 0.33$\,\dmunit) reporting consistent measurements. More recently, \cite{zhang_2025_atel} report a further decrease to $543.5 \pm 0.1$\,\dmunit\ in April 2025. Assuming a linear increase in the DM of \rmkt with time, this corresponds to $2.46$\,\dmunityr. This rate is approximately a factor of three higher than that of \rone during the short $3$\,month time span over which it was observed to have an increasing DM. In contrast, the DM may not be increasing linearly on longer timescales but could instead be fluctuating, and we may currently be observing an upward fluctuation. Both scenarios suggest that \rmkt, like \rone, is embedded in a dense magnetospheric environment.

Additionally, we measure a RM increase of $\sim$$80$\,\rmunit ($>40$\,\% fractional change) for \rmkt, see Section~\ref{sec:nrt-method}. Long-term monitoring of repeating FRBs also allows for the study of RM evolution and stability \citep{mckinven_2023_apj_repeaters, ng_2025_apj}. RM evolution is quite diverse among repeating FRBs. For \rone, long-term monitoring revealed a $\sim$$75\,\%$ decrease, from $\sim$$181,000$\,\rmunit to $\sim$$41,000$\,\rmunit, in nearly a decade \citep[RM values are quoted in the source rest frame,][]{michilli_2018_natur, plavin_2022_mnras, wang_2025_arxiv}. Likewise, \ronetwin shows RM variations on the order of $\sim$$20,000$\,\rmunit, including sign flips, leading to the hypothesis that the bursts are propagating through the stellar wind of a binary companion \citep{annathomas_2023_sci}. Aside from gradual changes, a rapid twentyfold RM increase and subsequent decrease was reported for \ronezeronine within the span of weeks in 2023 (an `RM flare'; \citealt{li_2025_arxiv}). These findings demonstrate the extreme local magneto-ionic environments that these sources reside in. In contrast, the hyperactive repeating source \roneoneseven maintains a stable RM of roughly zero suggesting a less dense and/or less magnetised local environment \citep{feng_2024_apj}. Initially, the RM of \rthree has been reported to vary only stochastically by $\sim$$3$\,\% \citep{mckinven_2023_apj}, but subsequent observations from 2021 and 2022 with CHIME/FRB and LOFAR revealed a decrease in the absolute RM of $\sim$$60$\,\rmunit \citep[$\sim$$50$\,\% fractional change,][]{mckinven_2023_apj, gopinath_2024_mnras}. The increase of the RM of \rmkt presented in this work is consistent with a subclass of repeating FRB sources that show variability in RM.

\subsection{Apparent DM variability within a single burst}
\label{sect:dm_var_single}
Evidence for DM variations within single bursts has previously been inferred from component misalignments that appear when the burst is de-dispersed using a single DM \citep[e.g.,][]{faber_2024_apj, platts_2021_mnras}. Moreover, \cite{hewitt_2023_mnras} found that the microshots within larger burst envelopes were mutually consistent in DM, while some of the broader (millisecond-duration) components of the burst appeared over-dedispersed, by up to $0.3$\,\dmunit. Nonetheless, the drifting component in B26-NRT (Appendix~Figure~\ref{fig:nrt_dm_opt}) is a clear case of intra-burst DM variability. Within the context of magnetospheric models of FRB emission, we will consider two plausible explanations: multiple emission regions, or plasma lensing.
\par
First, it is plausible that the drifting component is emitted at a lower emission height and thus becomes slightly more dispersed as it escapes the magnetosphere --- though the degree of dispersion will depend on how relativistic the intervening free electrons are. Accelerated particles whose emission contributes to a single complex FRB may propagate along different magnetic field loops, which are aligned with the observer at different emission heights, resulting in slightly different dispersion measures. However, generally magnetospheric FRB emission models predict that emission closer to the surface should have a larger luminosity and higher central frequency on account of the larger magnetic field; e.g., this is true for pair production mechanisms \citep{wadiasingh_repeating_2019}, coherent curvature radiation \citep{cooper_2021_mnras}, and coherent inverse-Compton scattering \citep{zhang_2022_ic}. In this picture, the brighter (and narrower) burst component, which has a higher dispersion measure than the drifting component, should come from a large emission height. This is counter to most magnetospheric FRB models, disfavouring this explanation.
\par
Second, the drifting component could be the result of plasma lensing of radio waves by the local ionised media within the host galaxy \citep{cordes_2017_apj}, and possibly in the local environment. We consider this possibility, as the detected long-term RM and DM variations for \rmkt (Section~\ref{sect:enviro}) imply an active, magnetised environment, which is consistent with the presence of lensing plasma in the source vicinity. As radio waves encounter a plasma lense, photons propagation along different paths with different deflection angles, $\theta$. The total group delay then contains two frequency-dependent terms: the dispersive delay, $\tau_{\rm disp} \propto DM(\theta) \nu^{-2}$, which may vary as a function of $\theta$ due to differences in free electron density in the plasma; and a geometric delay $\tau_{\rm geo} \propto \nu^{-4}$ due to longer path length, where the frequency dependence arises due to the plasma refraction index \citep{er_2020}. A combination of these factors means plasma lensing could explain why the drifting component of the FRB accumulates a larger DM. However, \citealt{main_2018_natur} show in a study of giant pulses that radio emission can be amplified, by factors of a few to tens, through plasma lensing. Generically, the emission bandwidth of lensed pulses scales inversely with the magnification $\upmu$ such that $\frac{\delta \nu}{\nu} \propto \upmu^{-1}$ for an elongated lens (or $\propto \upmu^{-1/2}$ for an elliptical lens). While the drifting component is less luminous than other burst components and appears (notwithstanding the limited observing bandwidth) to occupy a similar spectral bandwidth, it is plausible the burst was intrinsically less luminous prior to lensing. Unfortunately, we do not record the voltage data for the NRT and the data from Westerbork lacks sensitivity and bandwidth, thereby preventing the use of the the proposed lensing tests described by \cite{kader_2024_phrvd}. Although no definitive determination can be made, on balance it appears plasma lensing of the drifting component is consistent with our observations.

\subsection{Constraining the magnetic field strength of the local environment}
\label{sect:enviro}
The coincident detection of temporal variations in DM and RM means that, under certain assumptions, the parallel magnetic field component $B_{\parallel}$ of the intervening material can be estimated. We take the RM changes as determined by the NRT campaign for the polarization calibration file of MJD~$60598$, as it shows the smallest RM variation and thus provides the most conservative magnetic field estimate (Appendix~\ref{app:pcm} and Extended~Table~\ref{tab:burst-properties-nrt}). The NRT RM measurements span MJD~$60497$--$60537$, over which an increase in the absolute RM of $82.9\pm41.3$\,\rmunit\ was measured. We then fit a linear slope to the three DM measurement points obtained between MJD~$60511$--$60537$ by Westerbork (Extended~Table~\ref{tab:burst-properties-hf}). Extrapolating this fit back to MJD~$60497$, the first epoch of the RM measurements, yields a DM value of $464.7$\,\dmunit (see Supplementary material). This implies a DM change of $0.38$\,\dmunit\ over MJD~$60497$--$60537$, where we neglect the subdominant DM uncertainty. Under the approximation of a uniform medium across a line-of-sight length scale such that $\int_0^L n_e(l),dl = n_e L$:

\begin{equation}
B_{\parallel,{\rm mG}} = 1.23\,\textrm{mG}\,\left( \frac{\Delta\textrm{RM}}{\textrm{rad\,m}^{-2}}\right) \left( \frac{\Delta\textrm{DM}}{\textrm{pc\,cm$^{-3}$}}\right)^{-1}\approx 0.27\pm0.13 \, {\rm mG}
\end{equation}

We note that this parallel magnetic field estimate is lower than the estimates for \rone ($0.6-2.4$\,mG; \citealt{michilli_2018_natur} and $2.2-2.6$\,mG; \citealt{wang_2025_arxiv}), \ronetwin ($3-6$\,mG; \citealt{annathomas_2023_sci} ) and \reighteen ($>1.3$\,mG; \citealt{moroianu_2026_apj}). If we further assume the intervening matter is spherical, the DM variation time scale $\delta t$ implies via causality an upper limit on the length scale of $L < \delta t c \approx 0.03 \, {\rm  pc} = 9.3 \times 10^{11} \, {\rm km}$. This corresponds to a lower limit of the number density of free electrons in the material of $n_e \gtrsim 10 \, {\rm cm^{-3}}$ to achieve the observed $\delta \rm DM$ (e.g., $n_{e,\rm min}\sim \delta DM_{\rm obs}/L_{\rm max}$). Following \cite{michilli_2018_natur}, we can parametrise $B_{\parallel}$ in terms of the thermal particle energy at a temperature $T$ such that:

\begin{equation}
B_{\parallel} = (\beta^{-1} 16 \pi n_{\rm e} k_{\rm B} T)^{1/2}
\label{eq:B_parallel_ne}
\end{equation}

\begin{figure}
    \centering
    \includegraphics[width=1\linewidth]{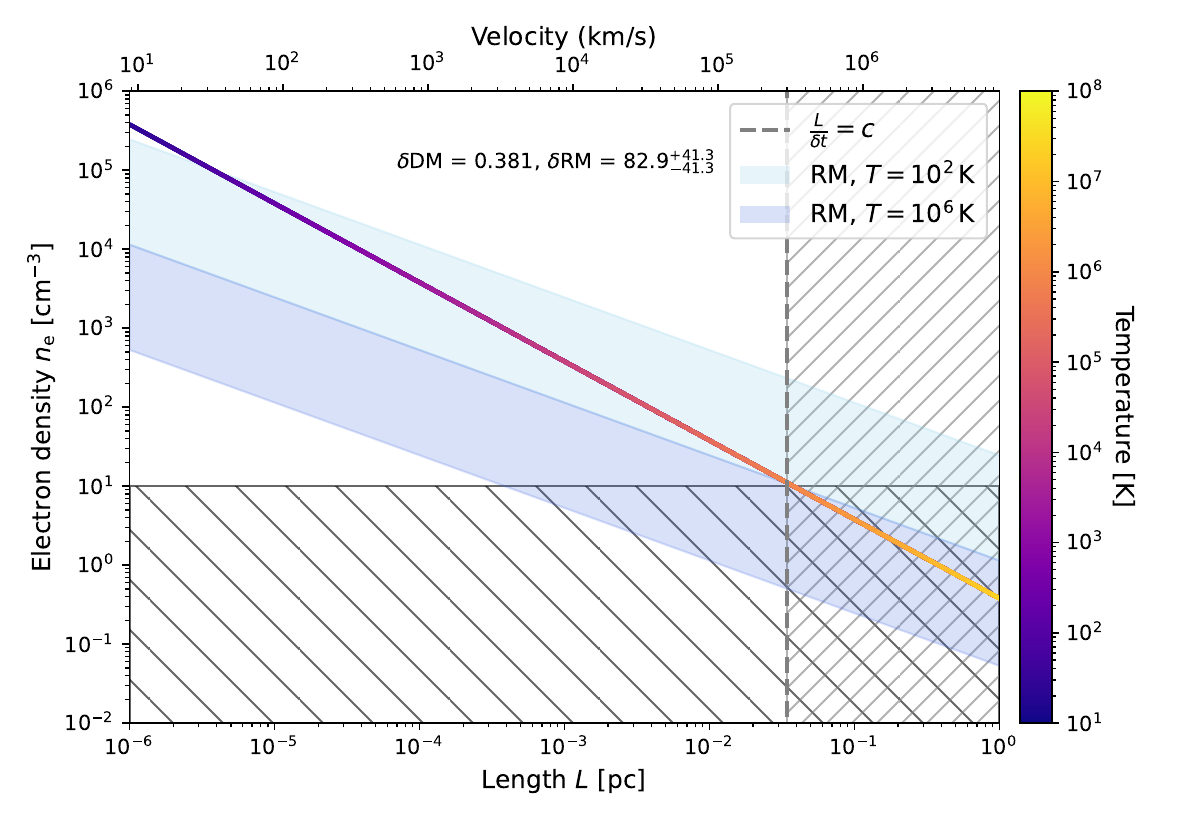}
    \caption{Constraints on the intervening matter using measured values of $\delta$DM and $\delta$RM via Equation~\ref{eq:B_parallel_ne}. The coloured line, and corresponding colour bar, denotes the required electron density $n_{\rm e}$ as a function of its longitudinal length scale $L$, coloured by the plasma temperature assuming equipartition (e.g., $\beta$ = 1). Background shades consider a wider allowed parameter space of $10^{-4} \leq \beta \leq 1$ for $T = 10^{2}$\,K (lighter shade) and $T = 10^{6}$\,K (darker shade). Cross-hatched areas correspond to intervening material velocities greater than the speed of light, assuming spherical geometry and $L \sim c \delta t$ where $\delta t$ is the observed timescale of RM and DM variations.}
    \label{fig:DM_RM_constraints}
\end{figure}

Where $\beta$ = 1 corresponds to equipartition between magnetic and particle energy densities. Using Equation~\ref{eq:B_parallel_ne}, we can express the required free electron density $n_{\rm e}$ in terms of $L$, $\beta$ and $T$. In Figure~\ref{fig:DM_RM_constraints} we show the constraints based on the observed RM variation for two plasma temperatures in background blue shades, which correspond to a range of $10^{-4} \leq \beta \leq 1$, where lower values of $\beta$ imply higher magnetizations. The causality constraints discussed above imply that the temperature of the free electrons must be $T\lesssim 10^{6} \, {\rm K}$ for $\beta = 1$, or $T \lesssim 10^{2} \, {\rm K}$ for $\beta = 10^{-4}$.

There are a number of plausible scenarios to simultaneously explain the DM and RM variations. Evolving supernova remnants (SNRs) have been posited to explain the RM and DM variations of some repeating FRBs (e.g., \citealt{yang_2017_apj}). While in late-time Sedov-Taylor phases increasing DM is predicted \citep{piro_2018_apj}, there is no prediction for an absolute RM increase. Furthermore, typical SNR length scales in this phase are larger than our inferred size constraints and should be slowly evolving, which is not supported by our observations. 
\par
Crab-like filaments have typical densities ($n_{\rm e} \sim 10^{3} \, {\rm cm^{-3}}$; \citealt{temim_2024}) and typical widths ($L\sim 10^{-4} - 10^{-2} \, {\rm pc}$; e.g., \citealt{hester_2008}) consistent with the allowed parameter space. Moreover, recent low-frequency radio observations \citep{aris2025} identified more diffuse envelopes with densities $n_{\rm e} \sim 10^{2} \, {\rm cm^{-3}}$ and typical widths of $L\sim 10^{-2} - 10^{-1} \, {\rm pc}$, also compatible with our constraints. However, while the magnitude of RM and DM changes could be explained, it is not clear that the velocities required to produce the rate of change are compatible with Crab-like filaments. Typical turbulent velocities of $v \sim 10^{2} \, {\rm km \, s^{-1}}$ \citep{Trimble_1971} appear incompatible with observations, and may required some fine-tuning: e.g., an observer aligned with non-uniform edge of a turbulent filament. 
\par
It has been suggested that some active repeaters may be magnetars in binary systems \citep{ioka_2020,wang_2022_bestar,zhang_2025_apjl}. The properties of a dense (possibly clumpy), warm, and magnetized stellar wind from a massive star are consistent with the constraints presented in Figure~\ref{fig:DM_RM_constraints} (see also predictions in \citealt{jakob_kaustubh}), and do not require large departures from equipartition. 
\par
Overall, the changing RM and DM appear to be best fit by either a clumpy stellar wind, or possibly filamentary structures surrounding the FRB source. Such structures may also give rise to the conditions required for plasma lensing discussed in Section~\ref{sect:dm_var_single}.

\subsection{Scintillation and Scattering} \label{sec:disc_scin_scat}

Our scattering analysis lends further evidence that there is a dense, variable magneto-ionic environment around the source. We find evidence for two distinct scattering media (hereafter `screens') along the line of sight. One screen produces a $\sim$$100$-kHz scintillation scale seen in seven out of nine of the bursts that were bright enough for the analysis. Given that pulsar observations indicate a spread of $\approx 0.8$ dex in Galactic scintillation bandwidths and pulse broadening times for a given DM \citep{krishnakumar15,cordes22}, we conclude that this $\sim100$ kHz scale is consistent with the the $\approx660$ kHz scale expected from the Milky Way foreground in this direction \citep{cordes_2002_arxiv}.

We additionally infer pulse broadening delays ranging between $120$\,$\upmu$s and $3$\,ms at $1$\,GHz. This level of pulse broadening is significantly larger than the level of scattering induced as $\sim$$100$\,kHz-scale scintillation, and we attribute it to an extragalactic screen. The detection of two scintillation scales, $\Delta \nu_{\rm d} = 156$\,kHz and $\Delta \nu_{\rm d} = 6.4$\,kHz at $1271$\,MHz, in burst B01-Wb (the brightest burst in our sample) further supports this conclusion. The extragalactic scintillation scale of $\Delta \nu_{\rm d} = 6.4$\,kHz corresponds to a scattering delay of $\tau \approx 1/(2\pi \ 6.4\, {\rm kHz}) \approx 24$\,$\upmu$s at 1271\,MHz and is consistent with the inferred upper limit on the scattering delay of $\tau_{\textrm{B01-Wb}} = 108$\,$\upmu$s (See Figure \ref{fig:tau}). In this burst, the inferred modulation index $m \approx 0.6$ of the Galactic $\sim $$100$-kHz scale suggests that extragalactic pulse broadening suppresses the Galactic scintillation. 

The extragalactic scattering appears to vary significantly between bursts, supporting an origin close to the source. Based on both the pulse broadening measurements and the $\Delta \nu_{\rm d} \approx 6.4$\,kHz measurement in B01-Wb, we infer scattering delays that vary by at least a factor of 2, and up to a factor of 9, between bursts. This level of variability is far too large to be attributable to refractive scattering in the host ISM (e.g., Galactic pulsars show scattering delays that vary by $\sim$$10$\,\% on months-long timescales; \citealt{singha24,geiger25}). Adopting the same line of reasoning applied to the DM and RM variations suggests that the scattering variations occur on length scales $\delta t c < 4\times 10^{11}$\,km (for $\delta t \approx 15$\,days between bursts B01-Wb and B02-Wb).

While this FRB currently lacks a known host galaxy, we can further attempt to constrain the position of the extragalactic scattering screen based on the DM budget. We first adopt a conservative upper limit on the source redshift based on its DM, which gives $z < 0.6$ ($95\%$ confidence, \citealt{connor_2025_natas}), or $D_{\rm s} < 1.4$\,Gpc \citep{planckcollaboration_2020_aa}. Based on the detection of both extragalactic pulse broadening and Galactic scintillation, we find an upper limit on the product $L_X L_G$, where $L_X$ is the distance between the source and extragalactic screen, and $L_G$ is the distance between the observer and Galactic screen, following methods laid out in \cite{cordes_2019_araa}. For the smallest extragalactic scattering delay, in burst B01-Wb ($\tau \approx 24$\,$\upmu$s at 1271\,MHz), we find $L_X L_G < 2\times10^3$\,kpc$^2$. 

Adopting the largest measured scattering delay ($\tau \approx 1.3$\,ms at 1228\,MHz) yields a significantly smaller upper limit, $L_X L_G < 42$\,kpc$^2$. For typical Galactic screen separations $L_G \sim 1$\,kpc, both of these upper limits accommodate a screen within the host galaxy, although we emphasize that these upper limits are extremely conservative due to the unknown source redshift. Adopting the much smaller redshift upper limit inferred from the energy distribution (see Section~\ref{sec:dis_red}), $z \lesssim 0.24$, yields an even smaller range $L_X L_G \sim 14 - 660$\,kpc$^2$. Future identification of the host galaxy and source redshift will yield more stringent constraints on $L_X$, providing an independent test of the link between scattering and the circum-source medium, as well as enabling constraints on the emission region size \citep{nimmo_2025_natur}.

The scattering variability reported here is highly reminiscent of the scattering behaviour of \ronetwin \citep{ocker_2023_mnras}, as well as scattering variations seen in the Crab pulsar \citep{lyne75,backer2000,mckee2018} and the Galactic Centre magnetar \citep{spitler14,pearlman18}. All of these other sources are known or inferred to reside in regions of highly inhomogeneous and filamentary plasma; both the Crab pulsar and Galactic Centre magnetar are young sources that are viewed through their natal circum-source material. Similar to \ronetwin, we find no clear evidence of a correlation between the DM/RM and $\tau$ variations, suggesting that these propagation effects arise from different (but potentially overlapping) regions of the circum-source environment. For \ronetwin, the timescale of the scattering variations is significantly faster than that of the DM and RM variations, potentially because the scattering arises from plasma inhomogeneities closer to the source \citep{ocker_2023_mnras}. For \rmkt, we are unable to distinguish between the timescales of the DM, RM, and scattering variations. Given that scattering arises from sub-au scale plasma density fluctuations, it is possible that the scattering we observe traces plasma fluctuations at much smaller spatial scales within the same evolving structure responsible for the secular change in DM and RM. The evidence for plasma lensing within a single burst, discussed in Section~\ref{sect:dm_var_single}, hints at a similar scenario. Indeed in the Crab pulsar, plasma lensing and variability in the pulse broadening delay are empirically connected \citep{backer2000,lyne2001,nadeau2024}. With enough burst detections, future studies of \rmkt may be able to draw similar connections between plasma lensing, pulse broadening variations, and underlying circum-stellar medium plasma properties (e.g., $\delta n_e$, size scale). 

\subsection{The burst fluence distribution} \label{sec:fluence_dist}

We fit separate power laws to the fluence distributions observed with NRT, Westerbork, and Stockert. The different slopes of these power-law fits suggest a break at approximately $\sim$$25$\,Jy\,ms in the cumulative fluence distribution, where the distribution flattens from a steeper power law with index $\gamma \sim -1.9$ to a flatter power law beyond the break with index $\gamma \sim -0.8$. We note that the burst rate observed with Stockert is lower than that measured with Westerbork. This difference can be attributed to the fact that the two telescopes did not observe strictly contemporaneously, and that Westerbork obtained nearly one and a half times more the total exposure during the activity window (see Appendix Figure \ref{fig:obs_campaign} and Table~\ref{tab:obs_coverage}). Additionally, FRB repeaters are known to exhibit highly variable burst rates during periods of activity, which makes it challenging to observe consistent rates in non-simultaneous observations \citep[e.g.,][]{konijn_2024_mnras}. Nevertheless, both telescopes accumulated sufficient exposure during the period in which \rmkt was active to probe and observe the flat tail of the cumulative distribution. The observation of a break in the distribution towards higher energies is not unique and has been found before in other very active repeaters, such as \rsixseven\ described by \cite{kirsten_2024_natas} and for \roneoneseven by \cite{ouldboukattine_2025_mnras}. In addition, \cite{huang_2025_raa} reported a tentative break in the cumulative energy distribution for \ronefourseven, \cite{hewitt_2022_mnras} for \rone and \cite{zhang_2024_natco} found a single outlier in the energy distribution for \meightone. 

In Figure~\ref{fig:cumulative_burstrate}, we highlight the brightest detection with NRT (B26-NRT) using a red circle (Figure~\ref{fig:b01_nrt_pol}). This burst was also detected using Westerbork (B01-Wb), see Figure~\ref{fig:wb_nrt_bright_burst}. We note that the inferred fluences of the bursts differ by a factor of $1.75$, with $771$\,Jy\,ms for B01-Wb and $441$\,Jy\,ms for B26-NRT. This difference arises because the observing bands of Westerbork and NRT only partially overlap, specifically towards the bottom of the band where the brightness of the spectra of the burst is increasing. The bottom of the NRT band is at $1228$\,MHz, while Westerbork observes down to $1207$\,MHz. The bright NRT detection deviates from the main, steeper distribution fitted for NRT and appears to align more closely with the flatter distributions fitted for Westerbork and Stockert. We do not attempt to fit a broken power law based on this single outlier. Nevertheless, this suggests that with more observations and exposure hours, NRT could have probed the flattening of the high-energy tail, as it did during a longer observing campaign on \roneoneseven \citep{ouldboukattine_2025_mnras}.

The detection of a break in the cumulative burst fluence distribution for \rmkt\ marks the fourth such measurement reported in the literature. This strengthens the evidence that a break in the energy distribution is a common feature of hyperactive repeating FRB sources. Bursts that deviate from the steep ($\gamma\sim-2.0$) power law occur roughly every $20$--$100$\,h \citep{kirsten_2024_natas, ouldboukattine_2025_mnras, huang_2025_raa}, although this occurrence rate depends heavily on repeater activity, which is known to vary significantly with time, radio frequency, and between sources \citep{xu_2022_natur, konijn_2024_mnras}. Detecting a break towards higher energies therefore requires hundreds to thousands of hours of exposure using an instrument with a completion threshold lower than $\sim$$20$\,Jy\,ms. For hyperactive repeaters, the high burst rate facilitates the detection of the brightest events and enables probing the high-energy tail of the distribution.

In our previous work, we placed a limit on the maximum energetics of \roneoneseven \citep{ouldboukattine_2025_mnras}. In this work, we are not able to place a similar constraint because our sample contains too few burst detections. Although \rmkt appears to have been sufficiently hyperactive \citep{tian_2025_mnras, kumar_2024_atel} to probe the high-energy tail, the low elevation of the source, as seen from the Northern Hemisphere, limited our observing time to a maximum of $3$\,h per day when the source was above an elevation of $10$\,\degree. Until now, most hyperactive FRB sources have been detected by the CHIME/FRB telescope \citep{chime_2021_atel, mckinven_2022_atel, shin_2024_atel}. These sources are located at high declinations making them excellent targets for high-cadence monitoring using the HyperFlash project. \rmkt represents the first hyperactive source detected and located by the MeerKAT telescope. Current and future observational surveys of the southern sky could detect a new hyperactive repeaters. Obtaining hundreds to thousands of hours of exposure on these future hyperactive repeaters would then require a HyperFlash-like project utilizing telescopes located in the Southern Hemisphere.

\subsection{Constraining the redshift of \rmkt} \label{sec:dis_red}

Currently, \rmkt has no identified host galaxy or measured redshift. In Section~\ref{sec:disc_scin_scat} we adopted a conservative redshift estimate of $z < 0.6$ \citep{connor_2025_natas}, but we can constrain the redshift more tightly. To do so, we employ two independent methods. The first makes use of the Macquart (DM$_{\rm IGM}$-$z$) relation \citep{macquart_2020_natur} under conservative yet realistic assumptions, while the second, novel method relies on the observed break in the cumulative energy distribution.

First, we derive a distribution for the smooth intergalactic medium contribution (DM$_{\rm IGM}$) by decomposing the observed DM of \rmkt, DM$_{\rm obs} = 464.857$\,\dmunit, into its constituent parts:

\begin{equation} \label{eq:DM}
{\rm DM_{obs} = DM_{MW} + DM_{IGM}} (z) + \sum_i~\frac{\mathrm{DM}_{\mathrm{halo},i}}{1+z_i} + \frac{\mathrm{DM_{host}}}{1+z},
\end{equation}

Where $\rm DM_{MW}$ is the Galactic contribution (including the disk and halo); the contributions of intervening halos $\mathrm{DM}_{\mathrm{halo},i}$ are summed over; and $\rm DM_{host}$ represents the contribution from the host galaxy and local environment in the source rest frame. The MW~disk contribution can be estimated using the YMW16 and NE2001 electron density models \citep{yao_2017_apj,cordes_2002_arxiv,ocker_2024_rnaas}. For the FRB's Galactic coordinates $(l, b) = (15.4559\degr, -23.3294\degr)$, we find for the YMW16 model $\rm DM_{MW,disk} = 62.7$\,\dmunit while NE2001 predicts $\rm DM_{MW,disk} = 94.3$\,\dmunit. To be more conservative, we adopt a uniform prior centred on the lower estimate from YMW16, $\rm DM_{MW,disk} \sim \mathcal{U}(62.7 \pm 0.2\,DM_{\rm MW,disk})$\,\dmunit. The MW halo component is less certain; based on recent constraints on the circum-galactic medium \citep{cook_2023_apj} we assume $\rm DM_{MW,halo} \sim \mathcal{U}(10, 111)$\,\dmunit. Combining these components, we obtain a total MW contribution of $\rm DM_{MW} = DM_{MW,disk} + DM_{MW,halo}$. Since the redshift is unknown, we assume a conservative DM$_{\rm halo}$ contribution of $5$\,\dmunit from surveys of intervening halos, e.g., \citet{simha_2023_apj}; for the host contribution, we assume a fixed value of $\rm DM_{host} \approx 50$\,\dmunit \citep{arcus_2021_mnras}. Following the formalism of \citet{macquart_2020_natur}, the mean IGM contribution is:

\begin{equation}
\mathrm{DM_{IGM}} = \frac{3 c H_0 \Omega_b f_{\rm IGM}}{8 \pi G m_p} \int_0^z \frac{f_e(z')(1 + z')}{E(z')} \, dz',
\end{equation}

where $H_0 = 67.4$\,km\,s$^{-1}$\,Mpc$^{-1}$, $\Omega_b = 0.049$, $f_{\rm IGM} = 0.85$ and $f_e(z') \approx 7/8$ \citep{planckcollaboration_2020_aa, niu_2022_natur}. Rearranging this to solve for $z$, we place a $90\%$\,confidence upper limit of $z < 0.37$. 

A flattening in the cumulative energy distribution in the high-energy tail has been observed for multiple repeating FRB sources, see Section~\ref{sec:fluence_dist}). The observed flattening in the energy distribution was enabled by high source activity combined with ample observing time, ranging from hundreds to thousands of hours. If the break in the cumulative energy distribution is a universal feature of repeating sources, similarly long exposures on sufficiently active sources should allow probing of the high-energy tail. Suppose this break consistently occurs at a characteristic energy scale across different repeaters, then it could serve as a rough distance indicator, effectively functioning as a pseudo-standard candle. By rewriting the energy function from \citet{macquart_2018_mnras}, we find:

\begin{equation} \label{eq:break_est}
    \frac{D_L^2}{(1+z)^2} = \frac{E_{\textrm{break}}}{4\pi~\mathcal{F}_{\textrm{break}}}
\end{equation}

Here $E_{\textrm{break}}$ is the shared breakpoint in the energy distribution of repeating sources in erg\,Hz$^{-1}$, and $\mathcal{F}_{\textrm{break}}$ is the breakpoint in fluence in Jy\,ms. $D_L^2$ is the luminosity distance and $z$ the redshift. This expression can be numerically solved if one assumes a shared breakpoint in $E_{\textrm{break}}$ and a measured breakpoint in $\mathcal{F}_{\textrm{break}}$.

For \rmkt we observe a break in the cumulative energy distribution at $25$\,Jy\,ms (see Figure~\ref{fig:cumulative_burstrate}). The energy distributions for various repeaters flatten in the range of $24$--$73$\,Jy\,ms, which corresponds to $5$--$25 \times 10^{30}$\,erg\,Hz$^{-1}$ when converting to specific energy (erg\,Hz$^{-1}$). We take a conservative margin on these values, $1$ to $30\times 10^{30}$\,erg\,Hz$^{-1}$, and use these as lower and upper bounds for $E_{\textrm{break}}$. We next numerically solve Equation~\ref{eq:break_est} and estimate the redshift of \rmkt to be $z=0.042$--$0.240$ (Supplementary material). 

We find that the $90\%$ confidence upper limit from the Macquart (DM$_{\rm IGM}$-$z$) relation, $z < 0.37$, is consistent with our estimate of $z = 0.042$--$0.240$, which is based on the assumption that the break in the energy distribution is a common feature among repeating FRB sources (Table~\ref{tab:energy-breaks}. A future host galaxy association of \rmkt\ and a measurement of the corresponding redshift will provide a direct test of this prediction.

\begin{table}
\caption{Energy breaks for different repeaters}
\label{tab:energy-breaks}
\resizebox{\columnwidth}{!}{%
\begin{tabular}{l c c l}
\hline
\hline
Source         & E-break [erg/Hz] & F-break [Jy ms] & Reference \\ \midrule
\rsixseven     & $5.0-8.0~\times~10^{30}$ & 24 - 39                        & \cite{kirsten_2024_natas}        \\
\roneoneseven  & $3.2~\times~10^{30}$     & 24                & \cite{ouldboukattine_2025_mnras} \\
\ronefourseven & $1.0-2.5~\times~10^{31}$ & 30 - 73                        & \cite{huang_2025_raa}          \\ 
\rmkt          & -                        & $\sim 25$                & This work                        \\ \bottomrule
\end{tabular}%
}
\end{table}

\section{Conclusions} \label{sec:conclusion}
We conducted a high-cadence observing campaign targeting \rmkt using the HyperFlash (Westerbork, Dwingeloo and Stockert telescopes) and \'ECLAT (\nancay Radio Telescope) monitoring programs. In total, we obtained more than $500$\,h of observations over a $4$-month period and detected $217$ bursts, including $10$\,bursts with fluences exceeding $25$\,Jy~ms. Our main conclusions can be summarized as follows:

\begin{itemize}
    \item We find burst-to-burst DM and RM variations, up to $0.41$\,\dmunit over $2$~months and $80$\,\rmunit over $40$~days, respectively (Table~\ref{tab:rmkt-properties} and Figure~\ref{fig:dm_rm_over_time}). We detect DM variations using coherently dedispersed Westerbork voltage data that reveal short temporal features ($\leq100$\,$\upmu$s). RM variations are detectable using the large bandwidth and accurate polarization calibration of NRT.
    
    \item In addition to DM variations between bursts, we also detect apparent intra-burst DM variability between sub-bursts in the brightest burst of our sample (B26-NRT; Figure~\ref{fig:b01_nrt_pol}). We argue that this drifting component could be the result of plasma lensing in the source's local environment or variable emission heights (Section~\ref{sect:dm_var_single}). 
    
    \item Using the observed variation in DM and RM we measure the parallel magnetic field strength of the local environment, $B_\parallel =0.27\pm0.13$\,mG. This value is an order of magnitude lower compared to measurements of \rone ($0.6$--$2.4$\,mG), \ronetwin ($3$--$6$\,mG) and \reighteen ($>1.3$\,mG), but represents one of the few cases where a direct estimate of $B_\parallel$ is possible.
    
    \item We also use the Westerbork voltage data to measure the scintillation bandwidth and scattering timescales for $7$~bursts (Figures~\ref{fig:scint} and \ref{fig:tau} and Table~\ref{tab:scint}). We find find evidence for two distinct screens along the line of sight. The first screen is detectable for multiple bursts and produces $\sim$$100$\,kHz scintillation scales and is associated with the Milky Way. The second screen was found for a bright single burst (B01-Wb) and has a scintillation bandwidth of $6.8$\,kHz and is most likely located in the host galaxy or local environment of \rmkt. 

    \item The (time-variable) propagation effects observed towards \rmkt demonstrate that it resides in a dense magnetospheric environment similar to \rone and \ronetwin, but likely less extreme. Furthermore, we suggest that the observed changes in RM and DM could result due to a clumpy stellar wind from a binary companion or filamentary structures surrounding the host, which may also explain the origin of the hypothesised intra-burst plasma lensing.
    
    \item We observe a break in the cumulative fluence distribution at $\sim$$25$\,Jy~ms (Figure~\ref{fig:cumulative_burstrate}). Breaks in the energy distributions of repeating FRB sources have previously been found and described for \rsixseven \citep{kirsten_2024_natas}, \roneoneseven \citep {ouldboukattine_2025_mnras} and \ronefourseven \citep{huang_2025_raa}. Our detection of a break in the cumulative energy distribution of \rmkt further supports the conclusion that this behaviour is common to repeating FRB sources in general.
    
    \item We place a 90\% confidence upper limit of $z<0.37$ based on the Macquart (DM$_{\rm IGM}$-$z$) relation. Additionally, assuming that the break in the energy distribution is a universal feature among repeaters, and occurs around the same characteristic energy range of $1$--$30\times 10^{30}$\,erg\,Hz$^{-1}$, we estimate a redshift of $z = 0.042$-$0.240$. These limits and estimates are consistent. A future host galaxy association and redshift determination of \rmkt will provide a direct test of these constraints.
\end{itemize} 

The HyperFlash and \'ECLAT monitoring campaigns will continue observing both known and newly discovered FRB sources, for hundreds to thousands of hours per year, enabling deeper insight into burst energetics and long-term variability of propagation effects.

\section*{Acknowledgements}
{
We thank Reshma Anna-Thomas, Amanda Cook, Jakob Faber and Jacco Vink for insightful discussions.
We thank the directors and staff of the participating telescopes for allowing us to observe with their facilities. 
The AstroFlash research group at McGill University, University of Amsterdam, ASTRON, and JIVE is supported by: a Canada Excellence Research Chair in Transient Astrophysics (CERC-2022-00009); the European Research Council (ERC) under the European Union's Horizon 2020 research and innovation programme (`EuroFlash'; Grant agreement No. 101098079); and an NWO-Vici grant (`AstroFlash'; VI.C.192.045).
A.~J.~C acknowledges support from the Oxford Hintze Centre for Astrophysical Surveys which is funded through generous support from the Hintze Family Charitable Foundation.
S.K.O. is supported by the Brinson Foundation through the Brinson Prize Fellowship Program, and is a member of the NANOGrav Physics Frontiers Center (NSF award PHY-2020265).
K.~N. is an MIT Kavli Fellow.
Z.~P. is supported by an NWO Veni fellowship (VI.Veni.222.295)
We express our gratitude to the operators and observers of the Astropeiler Stockert telescope: Thomas Buchsteiner, Elke Fischer and Hans-Peter L\"oge.
This work makes use of data from the Westerbork Synthesis Radio Telescope and the Dwingeloo Radio Telescope, both owned by ASTRON. ASTRON, the Netherlands Institute for Radio Astronomy, is an institute of the Dutch Scientific Research Council NWO (Nederlandse Oranisatie voor Wetenschappelijk Onderzoek). We thank the Westerbork operators Richard Blaauw, Jurjen Sluman and Henk Mulder for scheduling and supporting observations. We express our gratitude to all volunteers running the Dwingeloo Radio Telescope.
The \nancay Radio Observatory is operated by the Paris Observatory, associated with the French {\it Centre National de la Recherche Scientifique} (CNRS). We acknowledge financial support from the {\it Programme National de Cosmologie et Galaxies} (PNCG) and {\it Programme National Hautes Energies} (PNHE) of INSU, CNRS, France.
}


\section*{Data Availability}

The data that support the plots within this paper and other findings of this study are available under  \url{https://doi.org/10.5281/zenodo.18184163} or from the corresponding author upon reasonable request.
The scripts and Jupyter notebooks used to analyse the data, generate the plots and tables with the burst properties are available at \url{https://github.com/astroflash-frb/frb20240619d-ouldboukattine-2026}. \\
This work made use of the following software packages: \texttt{astropy} \citep{astropy:2013, astropy:2018, astropy:2022}, \texttt{Jupyter} \citep{2007CSE.....9c..21P, kluyver2016jupyter}, \texttt{matplotlib} \citep{Hunter:2007}, \texttt{numpy} \citep{numpy}, \texttt{pandas} \citep{mckinney-proc-scipy-2010, pandas_15831829}, \texttt{python} \citep{python}, \texttt{scipy} \citep{2020SciPy-NMeth, scipy_15716342}, and \texttt{tqdm} \citep{tqdm_14231923}.
This research has made use of NASA's Astrophysics Data System. 
Software citation information aggregated using \texttt{\href{https://www.tomwagg.com/software-citation-station/}{The Software Citation Station}} \citep{software-citation-station-paper, software-citation-station-zenodo}.

The FRB software pipeline written to process and search the baseband data can be found at \url{https://github.com/pharaofranz/frb-baseband}. 
{\tt jive5ab} can be found on \url{https://github.com/jive-vlbi/jive5ab}, {\tt Heimdall} is hosted at \url{https://sourceforge.net/projects/heimdall-astro/} and \texttt{FETCH} can be found at \url{https://github.com/devanshkv/fetch}. 
The pulsar package {\tt DSPSR} is hosted at \url{https://sourceforge.net/projects/dspsr/} and {\tt SIGPROC} can be retrieved from \url{https://github.com/SixByNine/sigproc}.


\newpage



\newpage

\appendix

\section{Observational campaign}

An overview of the observational campaign is given in Figure~\ref{fig:obs_campaign}. Additionally, the logs of the observations is provided in \texttt{.csv} format in the Supplementary Material. 

\section{Dynamic spectra of Westerbork bursts}

The time series and dynamic spectra of a subsect of bursts detected by Westerbork are shown in Figure~\ref{fig:wb_bb_family}.

\begin{figure*}
    \centering
    \includegraphics[width=0.9\textwidth]{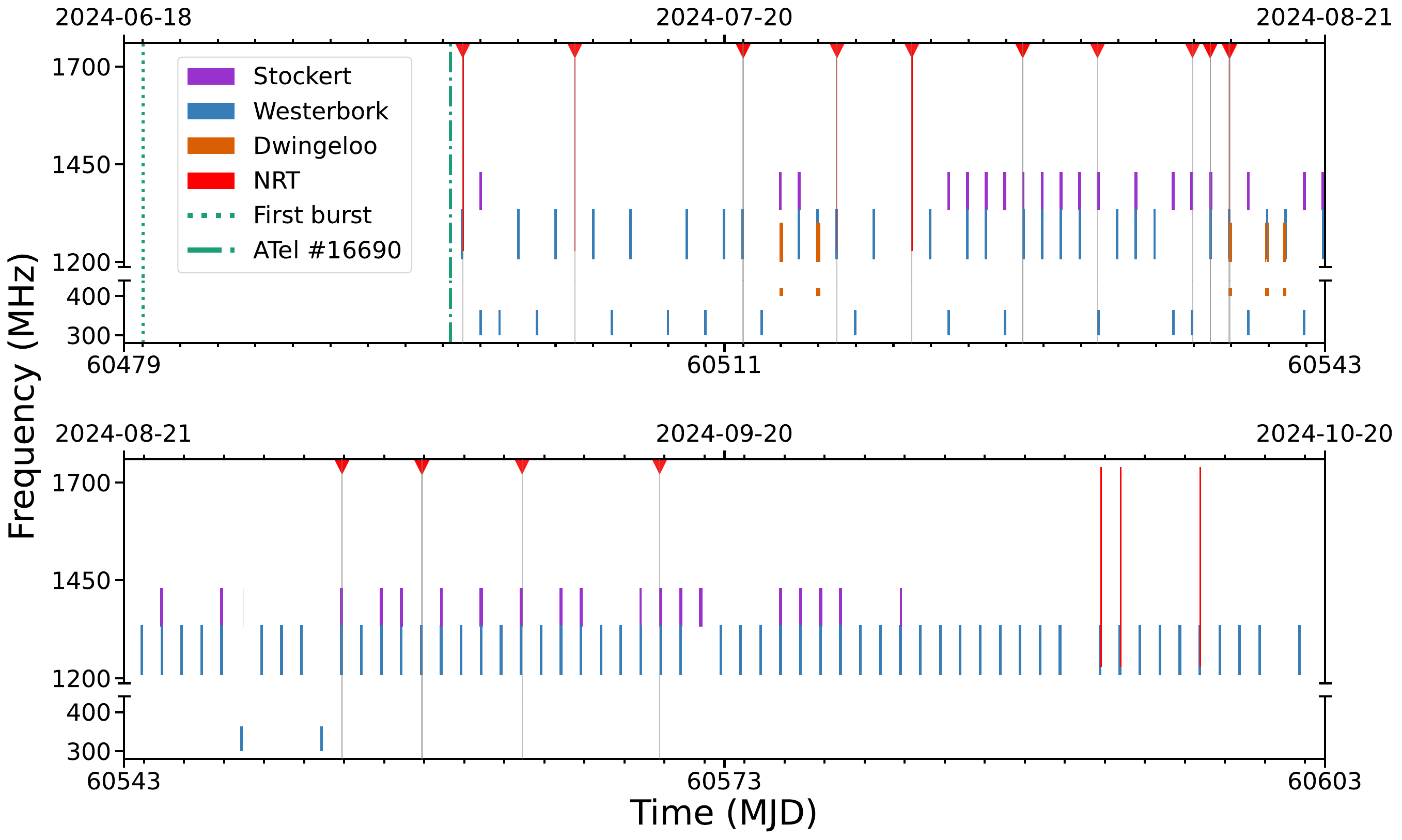}
    \caption{Overview of the observing campaign targeting \rmkt. Each coloured block represents an observation conducted with a specific telescope at a given frequency, where the block height is proportional to the observing bandwidth and the block width reflects the observation duration. The x-axis shows time in both MJD and calendar date; the top panel spans 64 days and the bottom panel 60 days. The broken y-axis indicates the frequency range of the observations. Red triangles, together with black vertical lines, mark the observations during which bursts were detected. In the top left of the plot, a dotted green line indicates the time of the first burst detected with MeerKAT, and a dash-dotted line marks the publication of the discovery ATel. Our observing campaign began 13 hours after the ATel was published.}
    \label{fig:obs_campaign}
\end{figure*}

\begin{figure*}
    \centering
    \includegraphics[width=0.9\textwidth]{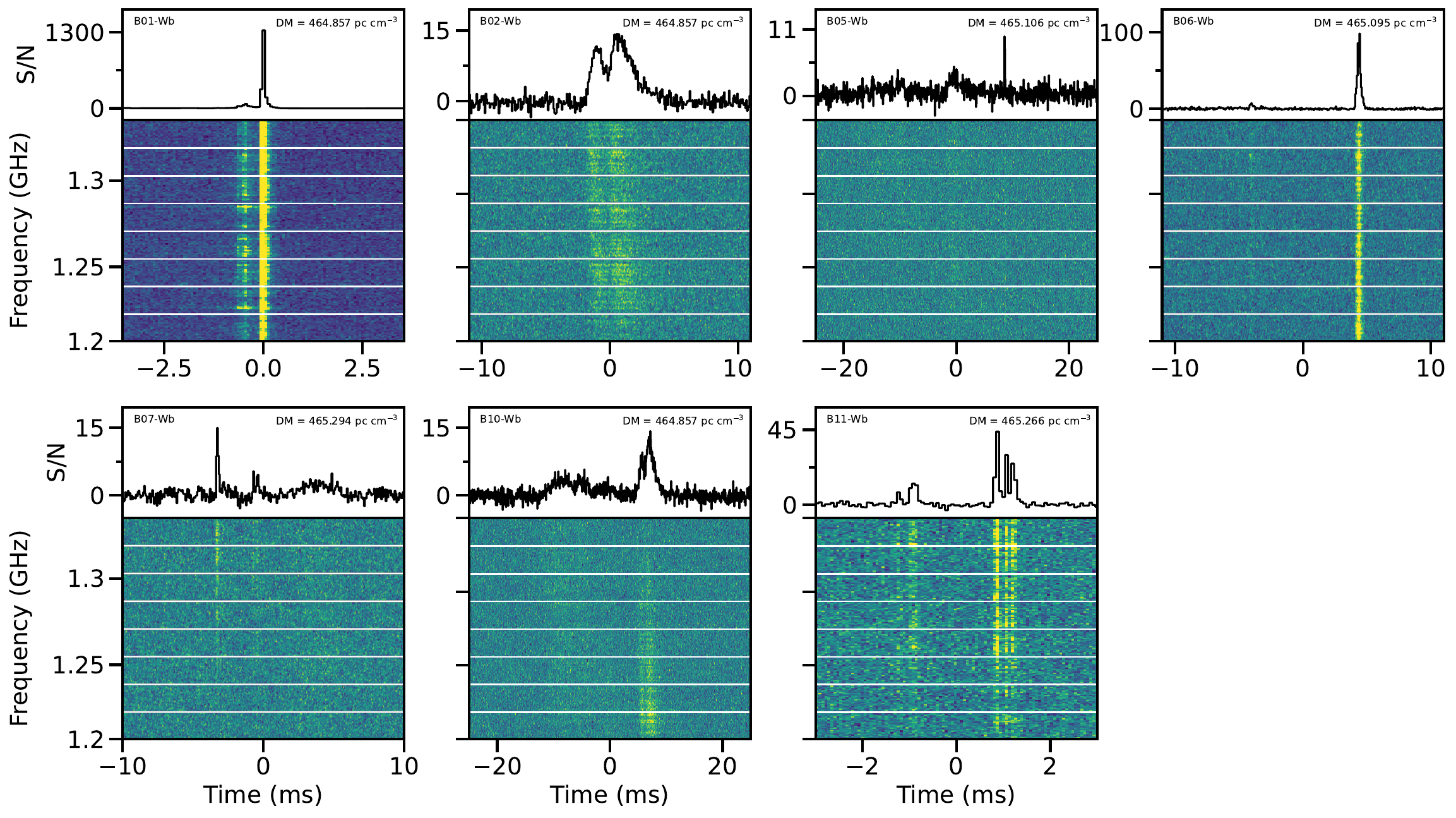}
    \caption{Time series (top panels) and dynamic spectra (bottom panels) for a subset of bursts that were detected by Westerbork, and for which voltage data was saved. The time and frequency resolution of all bursts shown here is $64\,\upmu \textrm{s}$ and $500\,$kHz, respectively. The bursts have been coherently dedispersed using \sfxc to the indicated DM in the top right. The white lines are zapped channels at the edges of the eight subbands used in the recording of the data. This plot illustrates that \rmkt bursts show complex temporal structures ranging from microseconds (B01-Wb) to milliseconds (B02-Wb).}
    \label{fig:wb_bb_family}
\end{figure*}

\section{Polarization Calibration}
\label{app:pcm}
\begin{figure}
    \centering
    \includegraphics[width=\columnwidth]{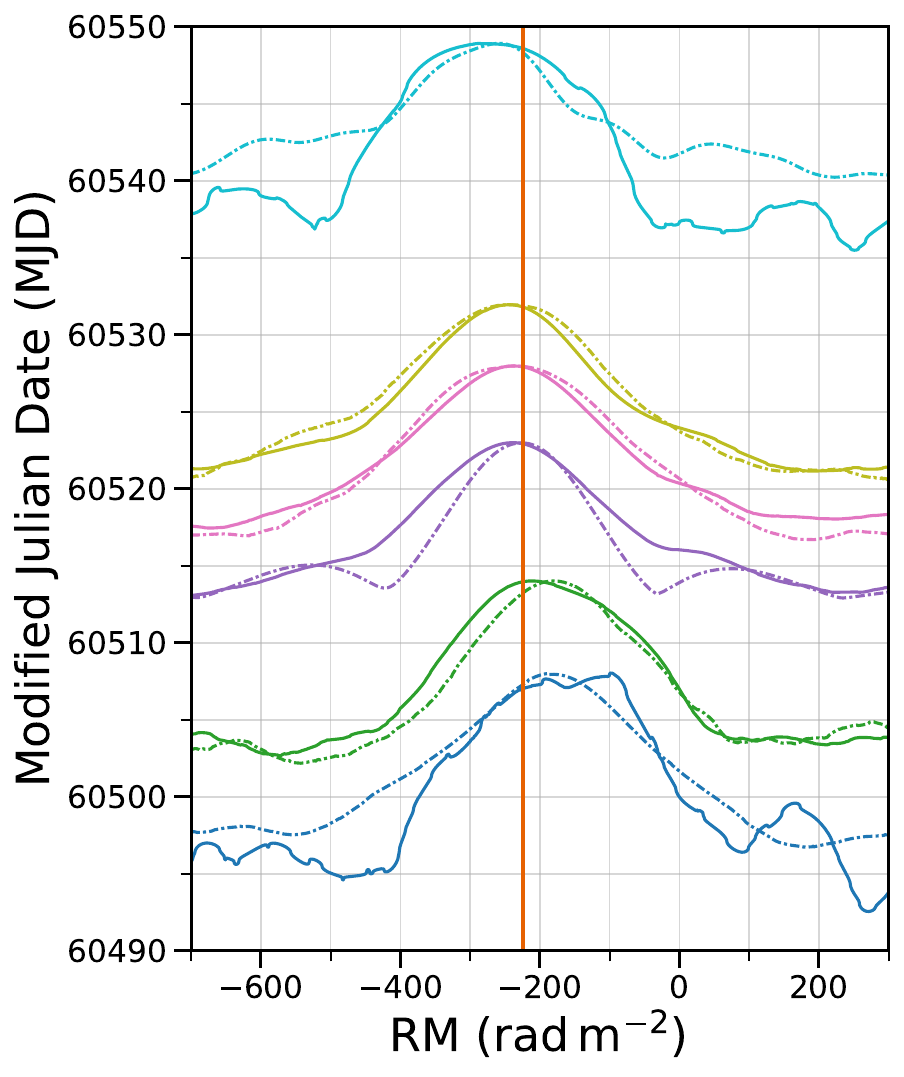}
    \caption{FDFs for the multiple \'ECLAT-detected bursts. The solid and dashes lines show the FDFs obtained from two different bursts selected from each observation and calibrated with the PCM file from MJD~60598. The vertical distance between FDFs corresponds to their time separation. More noisy FDFs correspond to weaker bursts. }
    \label{fig:JoyDivision}
\end{figure}

We calibrated the polarimetry of our NRT data with the full receptor model \citep{ord_2004_mnras,vanstraten_2004_apjs}, using both the noise diode scan from each observation and a polarization calibration modelling (PCM) file. By rotating the receiver horn by $180{\degree}$ over the course of a 1-hr observation of the pulsar PSR~J0742$-$2822, wide parallactic angle variations can be imitated, which in turn can be used to fully take into account the non-orthogonality of the receptor hands. This method has previously been used for polarisation calibration of pulsar data from the NRT \citep{guillemot_2023_aa}. Unfortunately, due to maintenance at the NRT, PCM files were only obtained on MJD~$60499$ and thereafter MJD~$60598$. We calibrated the data using both PCM files, ran \textrm{rmfit} to obtain a FDF, and fit a Gaussian to the central peak, shown in Figure~\ref{fig:JoyDivision}. Typically the RM values differ by $\lesssim6$\,\% between the two PCM files, and we thus conclude that the change in the absolute RM we see over the activity period (right panel in Figure~\ref{fig:dm_rm_over_time}) is not the result of inaccurate calibration.

\section{Burst property tables}

Burst properties from the HyperFlash and \'ECLAT observing campaigns are presented in Tables~\ref{tab:burst-properties-hf} and \ref{tab:burst-properties-nrt}. The full tables are provided in \texttt{.csv} format as part of the Supplementary Material and are also available on GitHub and Zenodo.

\begin{table*}
\caption{\textbf{Burst properties for the burst sample detected using HyperFlash. The full table is also available in \texttt{.csv} format in the Supplementary Material.}}
\label{tab:burst-properties-hf}
\resizebox{0.95\textwidth}{!}{%
\begin{tabular}{c c S[table-format=5.6,group-digits=none] S[table-format=3.2] S[table-format=4.2(5),separate-uncertainty=true] S[table-format=2.2] S[table-format=3.0] S[table-format=3.3(4),separate-uncertainty=true]}
\hline
\hline
{Burst ID} & {Station} & {TOA$^\mathrm{a}$} & {Peak S/N} & {Fluence$^\mathrm{b}$} & {Width} & {BW$^\mathrm{c}$} & {DM} \\
 &  & {[MJD]} &  & {[Jy ms]} & [ms] & [MHz] & {[pc cm$^{3}$]} \\ \midrule
B01 & wb & 60511.994427 & 750.47 & 771.73 \pm 154.35 & 27.71 & 128 & 464.857 \pm 0.003 \\
B02 & st & 60526.896118 & 17.98 & 100.75 \pm 20.15 & 6.99 & 98 & \multicolumn{1}{c}{--} \\
B02 & wb & 60526.896127 & 15.96 & 154.56 \pm 30.91 & 8.58 & 128 & \multicolumn{1}{c}{--} \\
B03 & st & 60530.882722 & 11.85 & 70.26 \pm 14.05 & 24.47 & 98 & \multicolumn{1}{c}{--} \\
B04 & st & 60535.945584 & 31.62 & 171.98 \pm 34.40 & 27.96 & 98 & \multicolumn{1}{c}{--} \\
B05 & st & 60536.884172 & 6.06 & 30.58 \pm 6.12 & 10.49 & 98 & \multicolumn{1}{c}{--} \\
B05 & wb & 60536.884182 & 10.10 & 102.92 \pm 20.58 & 25.47 & 128 & 465.106 \pm 0.013 \\
B06$^{\dagger}$ & dw & 60537.888446 & \multicolumn{1}{c}{--} & \multicolumn{1}{c}{--} & \sim1.00 & 200 & \multicolumn{1}{c}{--} \\
B06 & wb & 60537.888455 & 86.19 & 104.17 \pm 20.83 & 9.79 & 128 & 465.095 \pm 0.010 \\
B07 & st & 60553.898090 & 10.67 & 27.11 \pm 5.42 & 6.12 & 98 & \multicolumn{1}{c}{--} \\
B07 & wb & 60553.898099 & 16.28 & 41.32 \pm 8.26 & 10.30 & 128 & 465.294 \pm 0.079 \\
B08 & wb & 60557.865193 & 13.78 & 108.16 \pm 21.63 & 10.75 & 128 & \multicolumn{1}{c}{--} \\
B09 & wb & 60557.912320 & 9.85 & 8.10 \pm 1.62 & 3.84 & 128 & \multicolumn{1}{c}{--} \\
B10 & wb & 60562.892980 & 13.90 & 224.60 \pm 44.92 & 22.98 & 128 & \multicolumn{1}{c}{--} \\
B11 & wb & 60569.762476 & 35.32 & 47.13 \pm 9.43 & 3.52 & 128 & 465.266 \pm 0.016 \\
\bottomrule
\multicolumn{8}{l}{$\mathrm{^{\dagger}}$B06-dw is included, but fluence measurements were unavailable due to RFI; width and bandwidth were manually estimated.} \\
\multicolumn{8}{l}{$\mathrm{^{a}}$Time of arrival referenced to the solar system barycentre at infinite frequency in TDB.} \\
\multicolumn{8}{l}{\quad We adopt the DM of $464.857$ \dmunit for all bursts except for those for which we optimized the DM (Section \ref{sec:hf-method}).} \\
\multicolumn{8}{l}{\quad We assume a dispersion constant of $\mathcal{D} =1/(2.41 \times 10^{-4})$\,MHz$^{2}$\,pc$^{-1}$\,cm$^{3}$\,s.} \\
\multicolumn{8}{l}{\quad The position of Stockert is $X = 4031510.647$\,m, $Y = 475159.114$\,m and $Z = 4903597.840$\,m. The position of}\\
\multicolumn{8}{l}{\quad Westerbork is $X = 3828750.6969$\,m, $Y = 442589.2176$\,m and $Z = 5064921.5700$\,m; see the \href{https://github.com/jive-vlbi/sched/blob/python/catalogs/locations.dat}{EVN station locations} page.}\\
\multicolumn{8}{l}{$\mathrm{^{b}}$We assume a $20\%$ error for all bursts dominated by the uncertainty on the SEFD.} \\
\multicolumn{8}{l}{\quad For Westerbork the SEFD is retrieved from the \href{https://www.evlbi.org/sites/default/files/shared/EVNstatus.txt}{EVN status page}} \\
\multicolumn{8}{l}{$\mathrm{^{c}}$Measured bandwidth used to compute fluence.} \\
\end{tabular}%
}
\end{table*}

\begin{table*}
\caption{\textbf{Properties of the bursts detected using the \nancay Radio Telescope (NRT). The full table is available in \texttt{.csv} format in the Supplementary Material.}}
\label{tab:burst-properties-nrt}
\resizebox{0.95\textwidth}{!}{%
\begin{tabular}{
  c 
  S[table-format=5.6,group-digits=none] 
  S[table-format=1.2] 
  S[table-format=1.2(3),separate-uncertainty=true] 
  S[table-format=1.2] 
  S[table-format=3.0] 
  S[table-format=3.1(3),separate-uncertainty=true] 
  S[table-format=3.1(3),separate-uncertainty=true]
}
\toprule
\toprule
{Burst index} & {TOA$^\mathrm{a}$} & {Peak flux} & {Fluence$^\mathrm{b}$} & {Width$^\mathrm{c}$} & {Bandwidth$^\mathrm{c}$} & {RM$^\mathrm{d}$} & {RM$^\mathrm{d}$} \\
{[NRT]} & {[MJD]} & {[Jy]} & {[Jy ms]} & {[ms]} & {[MHz]} & {[rad m$^{-2}$]} & {[rad m$^{-2}$]} \\
\midrule
B01 & 60497.026715 & 0.62 & 0.76 \pm 0.15 & 0.76 & 144 & \multicolumn{1}{c}{--} & \multicolumn{1}{c}{--} \\
B02 & 60497.026983 & 1.63 & 3.89 \pm 0.78 & 3.88 & 188 & -182.2 \pm 31.9 & -187.4 \pm 28.2 \\
B03 & 60497.027644 & 0.82 & 2.36 \pm 0.47 & 2.36 & 432 & \multicolumn{1}{c}{--} & \multicolumn{1}{c}{--} \\
B04 & 60497.031234 & 0.62 & 0.98 \pm 0.12 & 0.98 & 428 & \multicolumn{1}{c}{--} & \multicolumn{1}{c}{--} \\
B05 & 60497.033447 & 0.73 & 1.24 \pm 0.25 & 1.24 & 188 & \multicolumn{1}{c}{--} & \multicolumn{1}{c}{--} \\
$\vdots$ & $\vdots$ & $\vdots$ & $\vdots$ & $\vdots$ & $\vdots$ & \multicolumn{1}{c}{\vdots} & \multicolumn{1}{c}{\vdots} \\
B202 & 60537.918644 & 0.80 & 3.37 \pm 0.67 & 3.37 & 184 & \multicolumn{1}{c}{--} & \multicolumn{1}{c}{--} \\
B203 & 60537.922310 & 1.20 & 0.76 \pm 0.15 & 0.76 & 184 & \multicolumn{1}{c}{--} & \multicolumn{1}{c}{--} \\
B204 & 60537.932108 & 0.72 & 1.40 \pm 0.28 & 1.40 & 112 & \multicolumn{1}{c}{--} & \multicolumn{1}{c}{--} \\
B205 & 60537.935503 & 3.05 & 1.81 \pm 0.36 & 1.80 & 232 & -254.9 \pm 43.1 & -272.4 \pm 30.6 \\
B206 & 60537.937718 & 0.72 & 1.01 \pm 0.20 & 1.00 & 184 & -279.8 \pm 46.9 & -270.3 \pm 30.2 \\
\bottomrule
\multicolumn{8}{l}{$\mathrm{^{a}}$Time of arrival referenced to the solar system barycentre at infinite frequency in TDB.} \\
\multicolumn{8}{l}{\quad We adopt a DM of $464.86$ \dmunit for all bursts in the NRT sample, assuming a dispersion constant} \\
\multicolumn{8}{l}{\quad of $\mathcal{D} =1/(2.41 \times 10^{-4})$\,MHz$^{2}$\,pc$^{-1}$\,cm$^{3}$\,s, reference frequency of $1738$\,MHz and source position} \\
\multicolumn{8}{l}{\quad as published in the discovery ATel of: \mktlocatel \citep{tian_2024_atel}.} \\
\multicolumn{8}{l}{\quad The position of NRT is: $X = 4324165.81$\,m, $Y = 165927.11$\,m and $Z = 4670132.83$\,m.}\\
\multicolumn{8}{l}{$\mathrm{^{b}}$We assume a $20\%$ error for all bursts, dominated by the uncertainty on the SEFD. } \\
\multicolumn{8}{l}{$\mathrm{^{c}}$Manually determined bandwidth and width used to compute fluence.} \\
\multicolumn{8}{l}{$\mathrm{^{d}}$Calibrated using the polarization calibration model (PCM) of MJD~60499 and 60598; see Appendix~\ref{app:pcm}.} \\
\end{tabular}%
}
\end{table*}

\section{Probing the narrowest frequency scale}

\begin{figure}
    \centering
    \includegraphics[width=\columnwidth]{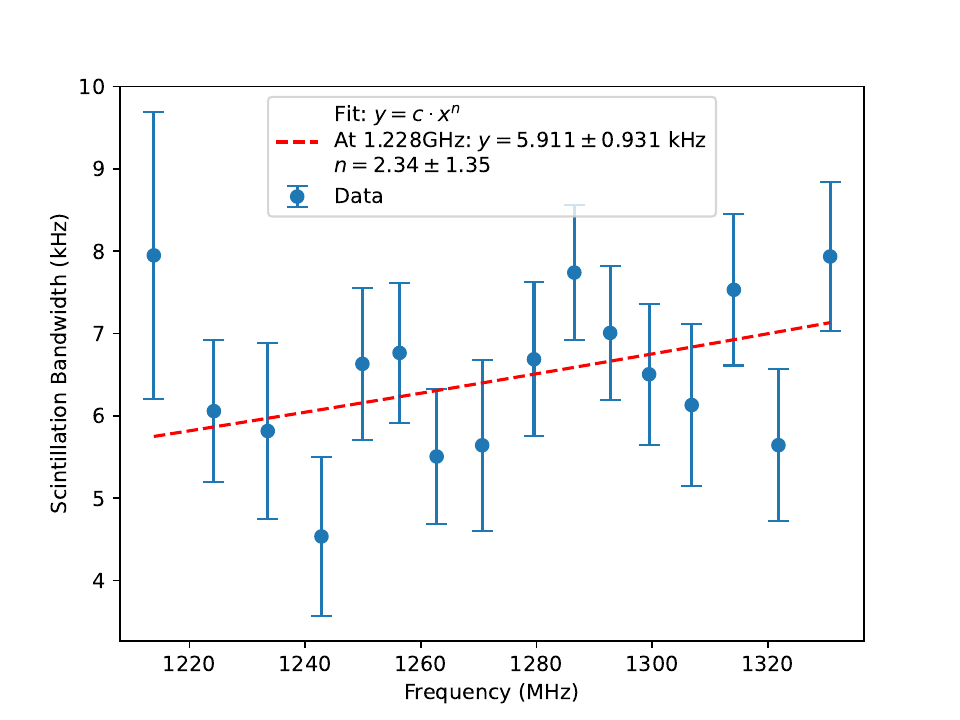}
    \caption{A measurement of the $\sim$$6$\,kHz scale across 16 subbands in the burst spectrum of B01-Wb. The frequency scale is measured across all subbands with a power-law frequency dependence of $\nu^{\alpha=2.34\pm1.35}$. }
    \label{fig:scin_freq_depen}
\end{figure}

In Figure~\ref{fig:scin_freq_depen} we show a measurement and tentative detection of the $\sim$$6$\,kHz scale for B01-Wb. 

\section{DM Optimisation for burst B26-NRT}

\begin{figure*}
    \centering
    \includegraphics[width=\textwidth]{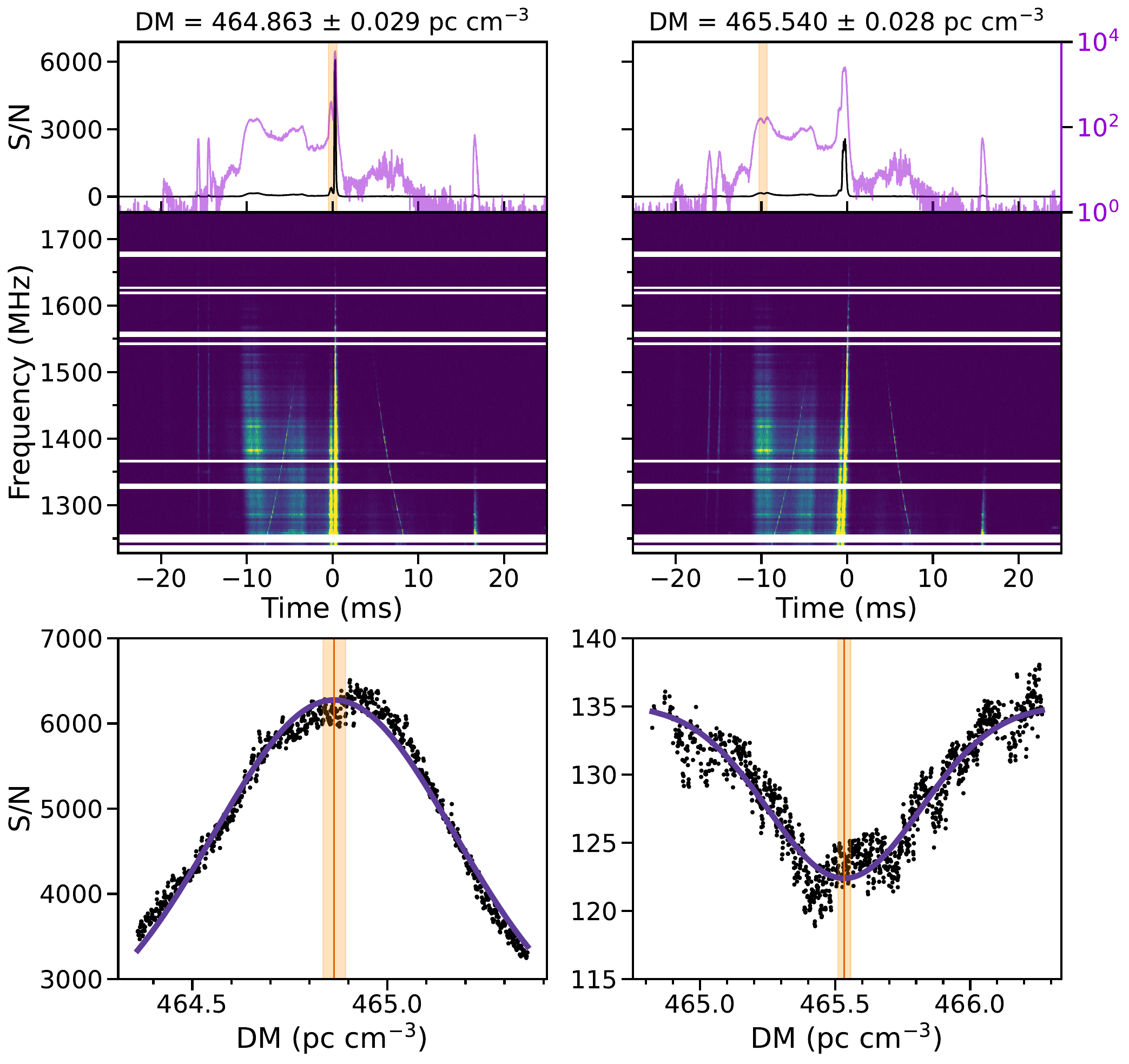}
    \caption{The brightest burst (B26-NRT) in the sample detected by NRT, dedispersed to two different DM values. The top panels display the dynamic spectrum and time series of the burst, with a yellow-highlighted region indicating the time window used to measure the peak S/N. We show the time series on both linear (black line) and logarithmic scales (purple line). The bottom panel shows the Gaussian fit: the solid orange line marks the best-fit value, and the yellow shaded region indicates the uncertainty on the DM.}
    \label{fig:nrt_dm_opt}
\end{figure*}

In Figure~\ref{fig:nrt_dm_opt} we show the DM optimisation of the bright burst detected, B26-NRT.

\bsp	
\label{lastpage}
\end{document}